
\input phyzzx
\hoffset=0.2truein
\voffset=0.1truein
\hsize=6truein
\def\TITLEPAGE{\frontpagetrue}
\def\CALT#1{\hbox to\hsize{\tenpoint \baselineskip=12pt
	\hfil\vtop{\hbox{\strut CALT-68-#1}
	\hbox{\strut DOE RESEARCH AND}
	\hbox{\strut DEVELOPMENT REPORT}}}}

\def\CALTECH{\smallskip
	\address{California Institute of Technology, Pasadena, CA 91125}}
\def\TITLE#1{\vskip 1in \centerline{\fourteenpoint #1}}
\def\AUTHOR#1{\vskip .5in \centerline{#1}}

\def\ABSTRACT#1{\vskip .5in \vfil \centerline{\twelvepoint \bf Abstract}
	#1 \vfil}
\def\ENDTITLEPAGE{\vfil\eject\pageno=1}

\def\sqr#1#2{{\vcenter{\hrule height.#2pt
      \hbox{\vrule width.#2pt height#1pt \kern#1pt
        \vrule width.#2pt}
      \hrule height.#2pt}}}

\def\section#1#2{
\noindent\hbox{\hbox{\bf #1}\hskip 10pt\vtop{\hsize=5in
\baselineskip=12pt \noindent \bf #2 \hfil}\hfil}
\medskip}

\def\underwig#1{	
	\setbox0=\hbox{\rm \strut}
	\hbox to 0pt{$#1$\hss} \lower \ht0 \hbox{\rm \char'176}}

\def\bunderwig#1{	
	\setbox0=\hbox{\rm \strut}
	\hbox to 1.5pt{$#1$\hss} \lower 12.8pt
	 \hbox{\seventeenrm \char'176}\hbox to 2pt{\hfil}}

\def \NP {{ Nucl. Phys. }}
\def \PL {{ Phys. Lett. }}

\def \PR {{ Phys. Rev. }}

\def \MPL {{ Mod. Phys. Lett. }}
\def \IJMP {{ Int. J. Mod. Phys.}}

\REF\Narain{K. S. Narain, \PL {\bf B169} (1986) 41.}
\REF\MM{M. Mueller, \NP {\bf B337} (1990) 37.}
\REF\V{G. Veneziano, \PL {\bf B265} (1991) 287.}
\REF\ATCOSMO{A. A. Tseytlin, Cambridge Univ. Preprints DAMTP-92-15 and
DAMTP-92-36; A. A. Tseytlin and C. Vafa, \NP {\bf B372} (1992) 443.}
\REF\MV{ K. A. Meissner and G.
Veneziano, \PL {\bf B267} (1991) 33 and \MPL {\bf A6} (1991) 3397; M.
Gasperini, J. Maharana, and G. Veneziano \PL {\bf B272} (1991) 277.}
\REF\GV{M. Gasperini and G. Veneziano, \PL {\bf B277} (1992) 256.}
\REF\CSF{E. Cremmer, J. Scherk, and S. Ferrara, \PL {\bf B74} (1978) 61.}
\REF\CJ{E. Cremmer and B. Julia, \NP {\bf B159} (1979) 141.}
\REF\CSS{E. Cremmer, J. Scherk, and J. H. Schwarz, \PL {\bf B84} (1979)
83.}
\REF\MS{N. Marcus and J. H. Schwarz, \NP {\bf B228} (1983) 145.}
\REF\JULIA{B. Julia, {\it in} ``Superspace and Supergravity,''
ed. S. Hawking and M. Ro\v cek
(Cambridge Univ. Press, Cambridge, 1980); P. Breitenlohner and D.
Maison, Ann. Inst. Poincar\'e {\bf 46} (1987) 215; H. Nicolai, \PL {\bf
B194} (1987) 402; H. Nicolai and N. P. Warner, Commun. Math. Phys. {\bf
125} (1989) 384.}
\REF\CHAM{A. Chamseddine, \NP {\bf B185} (1981) 403.}
\REF\BRDV{E. Bergshoeff, M. de Roo, B. de Wit, and P. Van Nieuwenhuizen,
\NP {\bf B195} (1982) 97.}
\REF\CM{G. Chapline and N. Manton, \PL {\bf B120} (1983) 105.}
\REF\HS{S. F. Hassan and A. Sen, \NP {B375} (1992) 103;
A. Sen \PL {\bf B271} (1991) 295 and \PL {\bf B272} (1992) 34.}
\REF\NSW{K. S. Narain, M. H. Sarmadi, and E. Witten, \NP
{\bf B279} (1987) 369.}
\REF\SS{J. Scherk and J. H. Schwarz, \NP {\bf B153} (1979) 61.}
\REF\CREMMER{E. Cremmer, {\it in} Supergravity '81, ed. S. Ferrara and
J.G. Taylor (Cambridge Univ. Press, Cambridge, 1982).}
\REF\BUSCHER{T. Buscher, \PL {\bf B194} (1987) 59; \PL {\bf B201} (1988)
466.}
\REF\SMITH{E. Smith and J. Polchinski, \PL {\bf B263} (1991) 59;
A. A. Tseytlin, Mod. Phys. Lett. {\bf A6} (1991) 1721.}
\REF\GS{M. B. Green and J. H. Schwarz, \PL {\bf B149} (1984) 117.}
\REF\ED{E. Witten, \PL {\bf B149} (1984) 351.}
\REF\SW{A. Shapere and F. Wilczek, \NP {\bf B320} (1989) 669;
A. Giveon, E. Rabinovici, and G. Veneziano, \NP
{\bf B322} (1989) 167; A. Giveon, N. Malkin, and E. Rabinovici, \PL {\bf
B220} (1989) 551; W. Lerche, D. L\"ust, and N. P. Warner, \PL {\bf B231}
(1989) 417.}
\REF\DUFFB{M. Duff, \NP {\bf B235} (1990) 610.}
\REF\GR{A. Giveon and M. Ro\v cek, preprint IASSNS-HEP 91/84.}
\REF\NEPO{R. A. Nepomechie, \PL {\bf B171} (1986) 195; J. M. Rabin, \PL
{\bf B172} (1986) 333; A. Das, P. Panigrahi, and J. Maharana, \MPL {\bf
A3} (1988) 759.}
\REF\JHS{J. H. Schwarz, ``Spacetime Duality in String Theory," p. 69 in
{\it Elementary Particles and the Universe},
ed. J. H. Schwarz (Cambridge Univ. Press 1991).}
\REF\DUFFA{M. Duff, \PL {\bf B173} (1986) 289.}
\REF\CFG{S. Cecotti, S. Ferrara, and L. Girardello, \NP {\bf B308}
(1988) 436.}
\REF\MO{J. Molera and B. Ovrut, \PR {\bf D40} (1989) 1146.}
\REF\JM{J. Maharana, ``Duality and $O(d,d)$ Symmmetries in String
Theory," Caltech preprint CALT-68-1781, April 1992.}
\REF\KKK{S. P. Khastgir and A. Kumar, \MPL {\bf A6} (1991) 3365; S. Kar,
S. P. Khastgir, and A. Kumar, \MPL (in press).}
\REF\BIANCHI{M. Bianchi, G. Pradisi, and A. Sagnotti, \NP {\bf B376}
(1992) 365.}
\REF\GSB{M. B. Green, J. H. Schwarz, and L. Brink,
Nucl. Phys. {\bf B198} (1982) 474.}
\REF\KY{K. Kikkawa and M. Yamasaki, \PL {\bf 149B} (1984) 357;
N. Sakai and I. Senda, Prog. Theor. Phys. {\bf 75} (1984) 692.}
\REF\NSSW{V. P. Nair, A. Shapere, A. Strominger,
and F. Wilczek, \NP {\bf B287} (1987) 402.}
\REF\RV{M. Ro\v cek and E. Verlinde, \NP {\bf B373} (1992) 630.}
\REF\GZ{M. Gaillard and B. Zumino, \NP {\bf B193} (1981) 221.}
\REF\EDWARD{E. Witten, Phys. Rev. Lett. {\bf 61} (1988) 670.}
\REF\AT{A. Tseytlin, \PL {\bf B242} (1990) 163; \NP {\bf B350} (1991)
395; Phys. Rev. Lett. {\bf 66} (1991) 545.}
\REF\FRADKIN{E. S. Fradkin and A. A. Tseytlin, \PL {\bf B158} (1985)
316; \NP {\bf B261} (1985) 1.}
\REF\CFMP{C. G. Callan, D. Friedan, E. Martinec, and M.J. Perry, \NP
{\bf B262} (1985) 593.}
\REF\FT{E. S. Fradkin and A. A. Tseytlin, Ann. of Phys. {\bf 162}
(1985)31.}
\REF\PANVEL{J. Panvel, Liverpool Preprint LTH282 (April 1992).}
\REF\SEN{A. Sen, Tata Institute preprints TIFR-92-20 and TIFR-92-29.}
\REF\SEIBERG{N. Seiberg, \NP {\bf B303} (1988) 286.}
\REF\GIVEON{A. Giveon and D. J. Smit, \NP {\bf B349} (1991) 168.}
\REF\CHSW{P. Candelas, G. Horowitz, A. Strominger, and E. Witten, \NP
{\bf B258} (1985) 46.}
\REF\Xenia{P. Candelas and X. C. de la Ossa, \NP {\bf B355} (1991) 455.}
\REF\CFGB{S. Cecotti, S. Ferrara, and L. Girardello, \IJMP {\bf A4}
(1989) 2457.}
\REF\DKL{L. J. Dixon, V. S. Kaplunovsky, and J. Louis, \NP {\bf B329}
(1990) 27.}
\REF\CXGP{P. Candelas, X. de la Ossa, P. Green, and L. Parkes, \NP {\bf
B359} (1991) 21.}
\REF\FFS{S. Ferrara, P. Fr\`e, and P. Soriani, Preprint CERN-TH 6364 and
SISSA 5/92/EP (Jan. 1992).}
\REF\FERRARA{S. Ferrara, Mod. Phys. Lett. {\bf A6} (1991) 2175 and
references therein.}
\REF\HUET{M. Dine, P. Huet, and N. Seiberg, \NP {\bf B322} (1989) 301.}

\TITLEPAGE
\CALT{1790}
\TITLE {Noncompact Symmetries in String Theory\foot{Work supported
in part by the U.S. Dept. of Energy under Contract no. DEAC-03-81ER40050.}}
\AUTHOR{Jnanadeva Maharana\foot {Permanent address: Institute of
Physics, Bhubaneswar --752005, India.} and {John H. Schwarz}}
\CALTECH
\ABSTRACT{Noncompact groups, similar to those that appeared in various
supergravity theories in the 1970's, have been turning up in
recent studies of string theory.  First it was discovered that moduli spaces
of toroidal  compactification are given by noncompact groups modded
out by their maximal compact subgroups and discrete duality groups.
Then it was found that many other moduli spaces have analogous descriptions.
More recently, noncompact group symmetries have turned up in effective
actions used to study
string cosmology and other classical configurations.  This
paper explores these noncompact groups in the case of toroidal
compactification both from the viewpoint of low-energy effective field
theory, using the method of dimensional reduction, and from the viewpoint
of the string theory world sheet.  The conclusion is that all these
symmetries are intimately related.  In particular, we find that
Chern--Simons terms in the three-form field strength $H_{\mu\nu\rho}$
play a crucial role.}

\ENDTITLEPAGE
\eject %

\chapter{\bf Introduction}

The unexpected appearance of noncompact global symmetries was one of the
most intriguing discoveries to emerge from the study of supergravity
theories in the 1970's.  The example that attracted the most attention
at the time was the $E_{7, 7}$ symmetry (a noncompact variant of $E_7$)
of $N = 8, D = 4$ supergravity.  More recently, noncompact groups have
been found to play a significant role in string theory.  Narain's
analysis of the heterotic string with $d$ toroidally compactified
dimensions
[\Narain]
focussed attention on the group $O(d, d + 16)$.  He showed
that the coset space $O(d, d + 16)/O(d) \times O(d + 16)$ is essentially
the moduli space of inequivalent compactifications.  Analogous coset
spaces describe the moduli spaces of certain Calabi--Yau, orbifold, and other
string compactifications, as well.  While Narain's $O(d, d + 16)$ group
is certainly not an exact symmetry of the
compactified heterotic string theory,
the discrete subgroup $O(d, d + 16 , Z)$ apparently is.
(This is an example of a
``target space duality" group, which relates distinct geometries
corresponding to the same conformal field theory.)
In the last couple of years, motivated
by considerations of superstring cosmology, attention has turned to the
study of what happens when the compactification moduli are allowed to be
time-dependent.  Mueller has found solutions with ``rolling radii" and a
time-dependent dilaton [\MM]. Veneziano discovered
an inversion symmetry for the cosmological scale factor,
or ``scale factor duality," both for vacuum solutions and for the
motion of classical strings in cosmological backgrounds [\V].
Similar observations were made by Tseytlin [\ATCOSMO] for the case of
closed strings and compact target space. Scale factor duality was later
extended to a full continuous
$O(d, d)$ symmetry of time-dependent (but independent of $d$ space
dimensions) solutions to the low-energy
theory both in the absence [\MV] and in the presence of classical string
sources [\GV].
The purpose of this paper is to explore the relationships between these
various appearances of noncompact global symmetry groups.  We will find
that they are all very closely related and that Chern--Simons terms play
a significant role in the realization of the symmetry.

The first appearance of a noncompact symmetry was the discovery of a
global $SU(1, 1)$ invariance in an appropriate formulation of $N = 4 , ~ D
= 4$ supergravity
[\CSF].
The qualification ``appropriate formulation"
refers to the fact that duality transformations allow $n$-forms to be recast
as $(D - n - 2)$-forms in $D$ dimensions $(d\tilde A = * dA)$, interchanging
the role of Bianchi identities and equations of motion.  Only after
appropriate transformations is the full noncompact symmetry exhibited.
In the $SU(1, 1)$ theory there are two scalar fields, which parametrize
the coset space $SU(1, 1)/U(1)$.  A year later, Cremmer and Julia showed
that $N = 8 , ~ D = 4$ supergravity could be formulated with $E_{7, 7}$
symmetry
[\CJ].
In this case the 70 scalars parametrize the coset
$E_{7, 7}/SU(8)$.  In analogous manner the $N = 8, ~ D = 5$ theory was
found to have $E_{6, 6}$ symmetry, with the 42 scalars parametrizing
$E_{6, 6}/USp (8)$
[\CSS].
The largest known symmetry of this type occurs in the $N
= 16, ~ D = 3$ theory, which has $E_{8, 8}$ symmetry, with 128 scalars
parametrizing $E_{8, 8}/O(16)$
[\MS].
It appears to be a general feature that
the scalars parametrize $G/H$, where $H$ is the maximal compact subgroup of
$G$. Two-dimensional examples, in which $G$ and $H$ are both infinite,
have also been considered [\JULIA].

The examples listed above (except for the $SU(1, 1)$ case) all refer to
maximally supersymmetric theories.  If they have any connection to
string theory, it is with the type II superstring.  Although it may be
worthwhile to do so, that line of inquiry will not be pursued here.
Rather, we shall focus on theories with half as much supersymmetry $(N =
1 ~ {\rm in} ~ D = 10 ~ {\rm or}~ N = 4 ~ {\rm in } ~ D = 4)$.  These
should be relevant to heterotic string theories.  In [\CHAM]
it was shown that the $N = 1, ~ D + d = 10$ supergravity theory,
dimensionally reduced to
$D$ dimensions (by dropping the dependence of the fields on $d$
dimensions), has global $O(d, d)$ symmetry. One exception
occurs for $D = 3$, where duality transformations allow
the symmetry to be extended to $O(8, 8)$ [\MS].  Moreover,
when the original $N = 1 ~ D = 10$ theory has $n$ Abelian vector
supermultiplets in addition to the supergravity multiplet, the global
symmetry of the dimensionally-reduced theory becomes extended to $O(d,
d + n)$, except for $d = 7$, where one obtains $O(8, 8 + n)$ [\MS].

The coupling of $N = 1 ~ D = 10$ supergravity to vector supermultiplets
require the inclusion of a Chern--Simons term $(H = dB - \omega_3)$ in
order to achieve supersymmetry.  This was shown in the Abelian case by
Bergshoeff {\it et al.} [\BRDV]
and in the non-Abelian case by Chapline and Manton [\CM].
In this paper we will focus on the bosonic sector, which
can be formulated in any dimension.  In section 2 we show that
dimensional reduction from $D + d$ dimensions to $D$ dimensions gives
rise to a theory with global $O(d, d)$ symmetry when there are no
vector fields in $D + d$ dimensions.  In section 4 the addition of $n$
Abelian vector fields in $D + d$ dimensions is considered.  We show that
the dimensionally-reduced theory has $O(d, d + n)$ symmetry provided
that the Chern--Simons term (described above) is included.  Thus the
desirability of such terms is deduced from purely bosonic
considerations!

The $O(d, d)$ symmetric theories considered by Veneziano and
collaborators [\V] are special cases of the
theories derived here.  Hassan and Sen have considered the extension to
$ n \not= 0$ and arbitrary $D$
[\HS].
However, for their purposes only the $O(d) \times O(d + n)$ subgroup is of
interest.

In the older supergravity theories, discussed above, a beautiful
technique for formulating the $G/H$ theory was developed.  One starts
with a matrix $V_{iA}$ of scalar fields belonging to the adjoint
representation of $G$, which acts as a sort of ``vielbein."  The $i$
index runs over a representation of $G$ and the $A$ index over the
corresponding
(possibly reducible) representation of the subgroup $H$.  Then the theory is
formulated with global $G$ symmetry and an independent local $H$
symmetry.  The latter is implemented by introducing auxiliary gauge fields
for the group $H$, without any kinetic term.  These fields, which are
somewhat analogous to the spin connection in a first-order formulation
of general relativity, can be eliminated by solving their equations of
motion (algebraically) and substituting back in the action.  The local
$H$ symmetry, which still is present after this substitution,
can then be used to choose a gauge
in which the scalar fields belonging to the $H$ subgroup are set to zero.
In section 3 we carry out this procedure explicitly for the $O(d, d)$
symmetric theory and show that it gives the correct action for the
moduli fields.
The vector fields are shown to form a $2d$-dimensional
vector multiplet of $O(d, d)$.  It is a general feature of the
supergravity theories that all bosonic fields other than the scalars are
inert under the local $H$ symmetry.
To our surprise, we discovered a second
construction that linearizes the action of $G$, which is also presented
in section 3.

In section 5 we reconsider the noncompact symmetries from the viewpoint
of the world-sheet ($\sigma$-model) action.  The result of Narain,
Sarmadi, and Witten [\NSW] that
the moduli of toroidal compactification parametrize $O(d, d)/O(d) \times
O(d)$ is briefly reviewed, as is the argument that string corrections
break the $O(d, d)$ symmetry to the discrete $O(d, d, Z)$ subgroup.  By
introducing $d$ coordinates $(\tilde Y_{\alpha})$ that are dual to the
$d$ compact string coordinates $(Y^{\alpha})$, we are able to obtain a
set of $2d$ classical equations of motion that have manifest $O(d, d)$
symmetry. The equations of motion for the space-time embedding of the
string $X^{\mu}$ are also recast in an $O(d,d)$ symmetric form.
The symmetry is broken to $O(d, d, Z)$ by boundary
conditions.

\chapter{\bf Dimensional Reduction Gives O(d,d) Symmetry}

In the 1970's it was noted that noncompact global symmetries
are a generic feature of supergravity theories containing scalar
fields.  One of the useful techniques that was exploited in these
studies was the method of ``dimensional reduction."  In its simplest
form, this consists of considering a theory in a spacetime M$\times$K,
where M has $D$ dimensions and K has $d$ dimensions, and supposing that the
fields are independent of the coordinates $y^{\alpha}$ of K.  For this to be a
consistent procedure it is necessary that K-independent solutions be
able to solve the classical field equations.  Then one speaks of
``spontaneous compactification" (at least when K is compact).  In a
gravity theory this implies that K is flat, a torus for example.  Of
course, in recent times more interesting possibilities,
such as Calabi--Yau spaces, have
received a great deal of attention. In such a case, the analog of
dropping $y$ dependence is to truncate all fields to their zero modes on
$K$. Here we will only
consider flat K, though generalizations would clearly be
deserving of study.

Explicit formulas for dimensional reduction were given in a 1979 paper by
Jo\"el Scherk and JHS [\SS] and subsequently developed further by
Cremmer [\CREMMER].  The main purpose of [\SS] was to
introduce a ``generalized" method of dimensional reduction that could
give rise to massive fields in the $D$-dimensional theory starting from
massless ones in the ($D + d$)-dimensional theory.  That procedure will
not be utilized here. Rather we will
stick to the simplest case in which the fields are taken to be
independent of the K coordinates.  Our notation is as follows:  Local
coordinates of M are $x^{\mu} (\mu = 0 , 1 , ... , D - 1)$ and local
coordinates of K are $y^{\alpha} (\alpha = 1 , ... , d)$.  The tangent
space Lorentz metric has signature $( - + ... + )$, unlike [\SS],
which results in a number of sign changes in the formulas given there.
All fields in $D+d$ dimensions are written with hats on the fields and
the indices ($\hat\phi$, $\hat g_{\hat\mu \hat\nu}$, etc.). Quantities
without hats are reserved for $D$ dimensions.
Thus, for example, the Einstein action on
M$\times$K (with a dilaton field $\hat\phi$) is
$$S_{\hat g} = \int_M dx ~ \int_K dy~ \sqrt{- \hat g}~ e^{-\hat\phi}
\big [\hat R
(\hat g) + \hat g^{\hat \mu \hat \nu} \partial_{\hat \mu} \hat\phi
\partial_{\hat \nu} \hat\phi \big ]\eqn\sba$$
If K is assumed to be a torus we can choose the coordinates $y^{\alpha}$
to be periodic with unit periods, so that $\int_K d y = 1$.  The radii
and angles that characterize the torus are then encoded in the metric
tensor.  As usual, the strength of the gravitational interaction is
determined by the value of the dilaton field.

The formulas that follow
can be read off from [\SS], generalized to include the dilaton
field. In terms of a ($D+d$)-dimensional vielbein, we can use local Lorentz
invariance to choose a triangular parametrization
$$\hat e^{\hat r}_{\hat \mu} = \left(\matrix {e^r_{\mu} &
A^{(1)\beta}_{\mu} E^a_{\beta}\cr 0 & E^a_{\alpha}\cr}\right ) \quad
{\rm and} \quad
\hat e_{\hat r}^{\hat \mu} = \left(\matrix {e_r^{\mu} & -
e_r^{\nu}A^{(1)\alpha}_{\nu}\cr 0 & E_a^{\alpha}\cr}\right )\, .\eqn\sbb$$
The ``internal" metric is
$G_{\alpha \beta} = E^a_{\alpha} \delta_{ab} E^b_{\beta}$
and the ``spacetime" metric is
$g_{\mu \nu} = e^r_{\mu}\eta_{rs} e^s_{\nu}$. As usual, $G^{\alpha \beta}$ and
$g^{\mu \nu}$ represent inverses and are used to raise the
appropriate indices.
In terms of these quantities the complete ($D + d$)-dimensional metric is
$$\hat g_{\hat \mu \hat \nu} = \left (\matrix {g_{\mu \nu} +
A^{(1)\gamma}_{\mu} A^{(1)}_{\nu \gamma} &  A^{(1)}_{\mu \beta}\cr
A^{(1)}_{\nu \alpha} & G_{\alpha \beta}\cr}\right ) \ {\rm and}
\ \hat g^{\hat \mu \hat \nu} = \left ( \matrix {g^{\mu \nu} &- A^{(1)\mu
\beta}\cr -A^{(1) \nu \alpha} & G^{\alpha \beta} + A^{(1)\rho \alpha} A^{(1)
\beta}_{\rho} \cr } \right ) .\eqn\sbc$$
A convenient property of this parametrization is that
$$ \sqrt{- \hat g} = {\rm det} \hat e^{\hat r}_{\hat \mu} = {\rm det}
e^r_{\mu} {\rm det} E^a_{\alpha} = \sqrt{-g} \sqrt{ {\rm det} G} \, .
\eqn\sbdet $$
If all fields are assumed to be $y$ independent, one finds after a
tedious calculation
$$\eqalign {S_{\hat g} =& \int_M dx \sqrt {-g}~ e^{-\phi}
\bigg\{ R + g^{\mu \nu}
\partial_{\mu} \phi \partial_{\nu} \phi\cr
&+ {1 \over 4} g^{\mu \nu} \partial_{\mu} G_{\alpha \beta} \partial_{\nu}
G^{\alpha \beta} - {1 \over 4} g^{\mu \rho} g^{\nu \lambda} G_{\alpha
\beta} F^{(1)\alpha}_{\mu \nu} F^{(1) \beta}_{\rho \lambda}
\bigg \}\ ,\cr} \eqn\sbd$$
where we have introduced a shifted dilaton field [\BUSCHER,\SMITH,\V]
$$\phi = \hat\phi - {1 \over 2} {\rm log~ det}\, G_{\alpha \beta}\,
\eqn\sbe$$
and
$F_{\mu \nu}^{(1) \alpha} = \partial_{\mu} A_{\nu}^{(1) \alpha} -
\partial_{\nu} A_{\mu}^{(1) \alpha}\, . $

Another field that is of interest in string theory is a second-rank
antisymmetric tensor $\hat B_{\hat \mu \hat \nu}$ with field strength
$$\hat H_{\hat \mu \hat \nu \hat \rho} = \partial_{\hat \mu} \hat
B_{\hat \nu \hat \rho} + {\rm cyc.~ perms}\,\, .\eqn\sbf$$
The Chern--Simons terms that appear in superstring theory are not
present here since we are not including ($D + d$)-dimensional vector
fields (in this section).  The Lorentz Chern--Simons term [\GS] is of higher
order in derivatives than we are considering.  The action for the $\hat
B$ term is
$$S_{\hat B} = - {1 \over 12} \int_M dx \int_K dy ~ \sqrt{- \hat g}~ e^{-
\hat\phi} ~ \hat g^{\hat \mu \hat \mu '}~ \hat g^{\hat \nu \hat \nu '} ~
\hat g^{\hat \rho \hat \rho '} ~ \hat H_{\hat \mu \hat \nu \hat \rho} ~
\hat H_{\hat \mu' \hat \nu' \hat \rho'} \,\, .\eqn\sbg$$

Because of the structure of the inverse metric, a little thought is
required to organize the terms in the dimensional reduction of eq. \sbg\
in a useful form.  A systematic
procedure is to first convert $\hat H$ to tangent space indices $\hat
H_{\hat r \hat s \hat t}$ and then use $e^r_{\mu}$ and $E^a_{\alpha}$
to convert back to Greek indices. This procedure leads to the result
$$\eqalign {S_{\hat B} = &- \int_M dx \sqrt{- g} ~ e^{- \phi} \bigg \{
{1 \over 12} H_{\alpha \beta \gamma} H^{\alpha \beta \gamma} +
{1 \over 4} H_{\mu \alpha \beta} H^{\mu \alpha \beta}\cr &+
{1 \over 4} H_{\mu \nu \alpha} H^{\mu \nu \alpha} +
{1 \over 12} H_{\mu \nu \rho} H^{\mu \nu \rho}
\bigg \}\,\, .\cr}\eqn\sbh$$
Here $H_{\alpha\beta\gamma}=0$, since $\hat B_{\alpha\beta}
=B_{\alpha\beta}$ is $y$ independent. Also,
$$H_{\mu\alpha\beta}=e_{\mu}^r \hat e_r^{\hat\mu} \hat H_{\hat\mu
\alpha\beta} =\hat H_{\mu\alpha\beta} = \partial_{\mu}B_{\alpha\beta}\, .
\eqn\sbaa$$
Similarly,
$$\eqalign {H_{\mu \nu \alpha} =\ & e_{\mu}^r e_{\nu}^s \hat e_r^{\hat
\mu} \hat e_s^{\hat\nu} \hat H_{\hat\mu \hat\nu \alpha}\cr
=\ & \hat H_{\mu\nu\alpha} -
 A_{\mu}^{(1)\beta}\hat H_{\beta\nu\alpha}
- A_{\nu}^{(1)\beta}\hat H_{\mu\beta\alpha}\cr
 =\ & F^{(2)}_{\mu \nu \alpha} - B_{\alpha \beta}
F^{(1) \beta}_{\mu \nu}\,\,,\cr} \eqn\sbi$$
where we have used
$F^{(2)}_{\mu \nu \alpha} = \partial_{\mu} A^{(2)}_{\nu \alpha} -
\partial_{\nu} A^{(2)}_{\mu \alpha} $
and
$$A^{(2)}_{\mu \alpha} = \hat B_{\mu \alpha} + B_{\alpha \beta}
A^{(1) \beta}_{\mu}\,\,.\eqn\sbj$$
The gauge transformations of the vector fields are simply $\delta A^{(1)
\alpha}_{\mu} = \partial_{\mu} \Lambda^{(1)\alpha}$ and $\delta
A^{(2)}_{\mu \alpha} = \partial_{\mu} \Lambda^{(2)}_{\alpha}$, under
which $H_{\mu \nu \alpha}$ is invariant.

For $H_{\mu \nu \rho}$ one finds
$$\eqalign {H_{\mu \nu \rho} =\ &e_{\mu}^r e_{\nu}^s e_{\rho}^t \hat
e_r^{\hat\mu} \hat e_s^{\hat\nu} \hat e_t^{\hat\rho}
\hat H_{\hat\mu\hat\nu\hat\rho} \cr
=\ & \hat H_{\mu\nu\rho} -\big(A_{\mu}^{(1)\alpha} \hat H_{\alpha\nu\rho}
+ {\rm 2\ perms}\big) +\big( A_{\mu}^{(1)\alpha} A_{\nu}^{(1)\beta}
\hat H_{\alpha\beta\rho}+ {\rm 2\ perms}\big)\cr
=\ & \partial_\mu B_{\nu \rho} - {1 \over 2}\big (A^{(1)
\alpha}_{\mu} ~ F^{(2)}_{\nu \rho \alpha} + A^{(2)}_{\mu \alpha} F_{\nu
\rho}^{(1)\alpha} \big ) + {\rm cyc.~ perms.\, ,}\cr}\eqn\sbk$$
where
$$B_{\mu \nu} = \hat B_{\mu \nu} + {1 \over 2} A^{(1) \alpha}_{\mu}
A^{(2)}_{\nu \alpha} - {1 \over 2} A^{(1) \alpha}_{\nu} A^{(2)}_{\mu
\alpha} - A^{(1) \alpha}_{\mu} B_{\alpha \beta} A^{(1) \beta}_\nu \ .
\eqn\sbl$$
In this case gauge invariance of the last line in eq. \sbk\
requires that under the $\Lambda^{(1)}$ and
$\Lambda^{(2)}$ transformations
$$\delta B_{\mu \nu} =  {1 \over 2} \big ( \Lambda^{(1) \alpha}
F^{(2)}_{\mu \nu \alpha} + \Lambda^{(2)}_\alpha F^{(1) \alpha}_{\mu \nu}
\big ) \,\, .\eqn\sbm$$
The extra terms in $H_{\mu \nu \rho}$,
which have arisen as a consequence of the dimensional reduction,
are abelian Chern--Simons terms. Recall that the
requirement that $H$ is globally defined implies that $dH$ is exact and
hence that ${\rm tr} (R \wedge R)
- {\rm tr} (F \wedge F)$ is exact for the familiar
Chern--Simons terms of $N = 1$, $D = 10$ supersymmetric theories [\GS,\ED].
In the present case, similar reasoning yields the requirement that $F^{(1)
\alpha} \wedge F^{(2)}_{\alpha}$ be exact.  Again, this is a significant
restriction on possible background configurations.

To recapitulate,
the dimensionally reduced form of $S = S_{\hat g} + S_{\hat B}$
has been written in the form
$$S = \int_M dx \sqrt{- g} ~ e^{- \phi} {\cal L}\,\, .\eqn\sbn$$
For the factor ${\cal L}$ we have found ${\cal L} = {\cal L}_1 + {\cal
L}_2 + {\cal L}_3 + {\cal L}_4$, where
$$\eqalign {{\cal L}_1 &= R + g^{\mu \nu} \partial_\mu \phi \partial_\nu
\phi\cr
	{\cal L}_2 &= {1 \over 4} g^{\mu \nu} \big (\partial_{\mu}
	G_{\alpha
	\beta} \partial_{\nu} G^{\alpha \beta} - G^{\alpha \beta}
	G^{\gamma \delta} \partial_{\mu} B_{\alpha \gamma} \partial_{\nu}
	B_{\beta \delta} \big )\cr
 	{\cal L}_3 &= - {1 \over 4} g^{\mu \rho} g^{\nu \lambda} \big (
	G_{\alpha \beta} F^{(1) \alpha}_{\mu \nu} F^{(1) \beta}_{\rho \lambda}
	+ G^{\alpha \beta} H_{\mu \nu \alpha} H_{\rho \lambda \beta}\big)\cr
	{\cal L}_4 &= - {1 \over 12}
	H_{\mu \nu \rho} H^{\mu \nu \rho}\,.\cr}\eqn\sbo$$

We now claim that there is an $O(d, d)$ global symmetry that leaves each
of these four terms separately invariant.  The first term $({\cal L}_1)$
is trivially invariant since $g_{\mu \nu}$ and $\phi$ are.  It should be
noted, however, that the individual terms in $\phi = \hat\phi - {1 \over 2}
{\rm log ~ det}\, G_{\alpha \beta}$ are not invariant.

To investigate the invariance of ${\cal L}_2$ we first rewrite it, using
matrix notation, as
$${\cal L}_2 = {1 \over 4} {\rm tr} \big ( \partial_\mu G^{- 1} \partial^\mu
G + G^{- 1} \partial_\mu B G^{- 1} \partial^\mu B \big )\,\,.\eqn\sbp$$
Then we introduce two $2d \times 2d$ matrices, written in $d \times d$
blocks, as follows [\SW]:
$$M = \pmatrix {G^{-1} & -G^{-1} B\cr
BG^{-1} & G - BG^{-1} B\cr}\eqn\sbq$$
$$\eta =  \pmatrix {0 & 1\cr 1 & 0\cr}\,\, .\eqn\sbr$$
Since $\eta$ has $d$ eigenvalues $+1$ and $d$ eigenvalues $-1$, it is a
metric for the group $O(d,d)$ in a basis rotated from the one with a
diagonal metric. The diagonal form will be used briefly in the next
section. Next we note that $M \in O(d, d)$, since
$$M^T \eta M = \eta \,\, .\eqn\sbs$$
In fact, $M$ is a {\it symmetric} $O(d, d)$ matrix, which implies that
$$M^{-1} = \eta M \eta = \pmatrix {G - BG^{-1} B & BG^{-1}\cr - G^{-1} B &
G^{-1}\cr }\,\,.\eqn\sbt$$
It is now a simple exercise to verify that
$${\cal L}_2 = {1 \over 8} {\rm tr} (\partial_\mu M^{-1} \partial^\mu
M)\,\,.\eqn\sbu$$
Thus ${\cal L}_2$ is invariant under a global $O(d, d)$ transformation
$$M \rightarrow \Omega M \Omega^T \, , \eqn\sbv$$
where
$$\Omega^T \eta \Omega = \eta \,\,.\eqn\sbw$$
This transformation acts on $G$ and $B$ in a rather complicated
nonlinear way.  We will give a simple description of this action later.

One might be tempted to think that the symmetry is even larger, since
${\cal L}_2$ is formally invariant under $M \rightarrow AMA^T$ for any matrix
$A \in GL (2d)$.  However, $M$ is not an arbitrary symmetric matrix
(which would have $d(2d -1)$ parameters), but one which belongs to $O(d,
d)$ and has just $d^2$ parameters.  Thus $O(d, d)$ transformations are
the most general transformations that preserve the structure of $M$ and
can be realized as transformations of $G$ and $B$.

Next we consider the ${\cal L}_3$ term:
$$\eqalign {{\cal L}_3 &= - {1
\over 4} \big [ F^{(1) \alpha}_{\mu \nu} G_{\alpha \beta} F^{(1)\mu \nu
\beta} + \big (F^{(2)}_{\mu \nu \alpha} - B_{\alpha \gamma} F^{(1)
\gamma}_{\mu \nu} \big ) G^{\alpha \beta} \big(F^{(2) \mu \nu}_\beta -
B_{\beta \delta} F^{(1) \mu \nu \delta} \big) \big]\cr &= - {1 \over 4}
{\cal F}^i_{\mu \nu} (M^{-1})_{ij} {\cal F}^{\mu \nu j} \,\,
,\cr}\eqn\sbx$$
where ${\cal F}^i_{\mu \nu}$ is the $2d$-component vector of field
strengths
$${\cal F}^i_{\mu \nu} = \pmatrix {F^{(1) \alpha}_{\mu \nu}
\cr F^{(2)}_{\mu \nu \alpha}\cr} = \partial_\mu {\cal A}^i_\nu - \partial_\nu
{\cal A}^i_\mu \,\, .\eqn\sby$$
Then ${\cal L}_3$ is seen to be $O(d, d)$ invariant
provided that the vector
fields transform linearly according to the vector representation of
$O(d, d)$, i.e., ${\cal A}_{\mu}^i \rightarrow \Omega^i{}_j {\cal A}^j_\mu $.

Finally we turn to ${\cal L}_4$.  In this case
$H_{\mu \nu \rho}$ can be written in the form
$$H_{\mu \nu \rho} = \partial_\mu B_{\nu \rho} - {1 \over 2} {\cal A}^i_\mu
\eta_{ij} {\cal F}^j_{\nu \rho} + ({\rm cyc.~ perms.})\,\,.\eqn\sbz$$
This is $O(d, d)$ invariant if we require that $B_{\nu \rho}$ not
transform. The second term is invariant since $\Omega^T
\eta \Omega = \eta $.

To make contact with string theory, the formulas we have presented here
are appropriate to the massless fields of the closed oriented bosonic
string with $D + d = 26$.  In that case the $O(d, d)$ symmetry is
certainly broken by higher mass and higher dimension terms that have
been dropped.  An $O(d, d, {Z})$ subgroup is believed to survive as
an exact symmetry of the theory, though it is broken spontaneously when
a particular background is selected.  This discrete group and its
relationship to the continuous groups described here will be explored in
section 5. To make contact with the heterotic string,
Yang--Mills gauge fields should be introduced in the original $(D +
d)$-dimensional theory.  This extension will be
explored in section 4.

\chapter{\bf Coset Space Reformulations}

The realization of $O(d, d)$ symmetry found in the last section, $M
\rightarrow \Omega M \Omega^T$, is not very transparent as a rule for
the transformation of the $d^2$ scalar fields
$$X = G + B\, .\eqn\sca$$
Let us explore this in a little detail.  To start with, consider an
infinitesimal $O(d, d)$ transformation given by
\foot{The $GL(d,R)$ subalgebra parametrized by the matrix $\alpha$
corresponds to constant (global) general coordinate transformations of
the internal manifold $K$. Clearly, in view of the toroidal topology,
only the $SL(d,Z)$ subgroup belongs to Diff($K$) (see section 5). The
remaining generators of $O(d,d,Z)$ correspond to integer shifts of the
moduli $B_{\alpha \beta}$.}
$$\Omega = \pmatrix {1 + \alpha & \beta\cr
	\gamma & 1 - \alpha^T\cr}\, , \eqn\scb$$
where $\alpha , \beta , \gamma$ are infinitesimal $d \times d$
matrices and $\beta = - \beta^T$, $\gamma = - \gamma^T$.  Then
$$\delta M = \pmatrix {\alpha & \beta \cr \gamma & - \alpha^T \cr}~ M +
M ~ \pmatrix {\alpha^T & - \gamma \cr - \beta & - \alpha \cr}\,
,\eqn\scc$$
which is easily seen to correspond to [\DUFFB]
$$\delta X = \gamma - \alpha^T X - X \alpha - X \beta X \, .\eqn\scd$$

The similarity of the last formula to one for an infinitesimal $SL (2 ,
C)$ transformation, which exponentiates to $z \rightarrow ( a\, z +
b) (c\, z + d)^{-1}$, suggests the following.  Write an arbitrary $O(d,
d)$ matrix $\Omega$ in block form
$$\Omega = \pmatrix { \Omega_{11} & \Omega_{12}\cr \Omega_{21} &
\Omega_{22}\cr}\, \, .\eqn\sce$$
Then the finite transformation
$$X \rightarrow (\Omega_{22} X + \Omega_{21}) (\Omega_{11} + \Omega_{12}
X)^{-1} \eqn\scee$$
reproduces the infinitesimal transformation formula obtained above [\GR].
Moreover, it has the correct group property, and so must be correct in
general.  The matrix that appears here is actually
$$\tilde \Omega = \eta \Omega \eta = \pmatrix {\Omega_{22} & \Omega_{21}
\cr \Omega_{12} & \Omega_{11}\cr} \, \, , \eqn\scf$$
which is an equivalence transformation.  It is the matrix
$X^{-1}$ that undergoes a linear fractional transformation controlled
by the matrix $\Omega$. This transformation law of $X$ is reminiscent
of that for
period matrices matrices under symplectic modular transformations in the
theory of Riemann surfaces.

How should we utilize these facts?  A possible goal is to rewrite the
action in terms of the matrix $X$ rather than the matrix $M$.  Another
possible goal is to introduce auxiliary gauge fields and extra scalar
fields such that the $O(d, d)$ symmetry is realized linearly.  Towards
these ends let us introduce a second real $d \times d$ matrix of scalar
fields, called $Y$, and generalize eq. \scee\ to
$$\pmatrix {X
\cr Y \cr}~ \rightarrow ~ \pmatrix {\Omega_{22} & \Omega_{21}\cr
\Omega_{12} & \Omega_{11} \cr} ~~ \pmatrix {X \cr Y \cr}\, \, ,\eqn\scg$$
which corresponds to the previous nonlinear transformation
rule for the matrix $XY^{-1}$.  In
other words, \scee\ corresponds to \scg\ in the ``gauge" $Y = 1$.  It
is convenient to introduce a $2 d \times d$ matrix $V$ consisting of the
blocks $X$ and $Y$ (as in eq. \scg), such that the
above transformation is
$$V_{i \alpha}
\rightarrow \tilde \Omega_{ij} ~ V_{j \alpha} \,\, .$$
The rectangular matrix $V_{i
\alpha}$ transforms as $d$ copies (labeled by $\alpha$) of the vector
representation of $O(d ,d)$.

In order to have enough gauge freedom to eliminate $Y$, which is an
arbitrary real nonsingular matrix, we need local
$GL (d , { R})$ gauge symmetry.  If $m_{\alpha \beta}$ is a matrix
belonging to $GL (d , { R})$, we require that $V_{i \alpha}$
transform as $2d$ copies (labeled by $i$) of the vector representation
of $GL (d , { R})$
$$V_{i \alpha} \rightarrow m_{\alpha \beta} ~ V_{i \beta} = (V m^T)_{i
\alpha} \,\, .\eqn\sch$$
Next we introduce auxiliary gauge fields, belonging to the $GL (d, {
R})$ algebra, called $(	A_{\mu})_{\alpha \beta}$, and we
define a covariant derivative
$$D_{\mu} V_{i \alpha} = \partial_{\mu} V_{i \alpha} +
(A_{\mu})_{\alpha \beta} V_{i \beta}\,\, .\eqn\sci$$

Now let us try to write a $V$ kinetic term with global $O(d, d)$ symmetry
and local $GL(d , { R})$ symmetry.  Two $O(d, d)$ invariant $d \times
d$ matrices are $(V^T \eta V)_{\alpha\beta}$ and
$(D_{\mu} V^T \eta ~ D^{\mu} V)_{\alpha\beta}$.  Under
local $GL (d, { R})$ transformations
$$\eqalign {V^T \eta V &\rightarrow m(V^T \eta V) m^T\cr
D_{\mu} V^T \eta D^{\mu} V &\rightarrow m(D_{\mu} V^T \eta D^{\mu}
V)m^T \, .\cr}\eqn\scj$$
Therefore the natural guess with the desired symmetries is
$${\cal L'}_2 = {1 \over 4} {\rm tr} \big[ (V^T \eta V)^{-1} (D_{\mu} V^T \eta
D^{\mu} V)\big]\,\,.\eqn\sck$$
It is straightforward to solve the classical field equation implied by
this Lagrangian for
$(A_{\mu})_{\alpha \beta}$ in the $Y = 1$ gauge with the result
$$(A_{\mu})_{\alpha \beta} = - {1 \over 2} (G^{-1})_{\beta \gamma}
\partial_{\mu} (G + B)_{\gamma \alpha}\,\,.\eqn\scl$$
Substituting this back into ${\cal L'}_2$, one obtains the desired result
found in Section 2:
$${\cal L'}_2 = {\cal L}_2 =
{1 \over 4} tr (\partial_{\mu} G^{-1} \partial^{\mu} G +
G^{-1} \partial_{\mu} B G^{-1} \partial^{\mu} B)\,\,.\eqn\scm$$

To complete this part of the story we still need to recast ${\cal L}_3$
in terms of $V$ in an $O(d, d) \times GL(d, { R})$ invariant form.
Since ${\cal F}^i$ is an $O(d, d)$ vector, $(V^T {\cal F})_{\beta}$ is
$O(d, d)$ invariant and a $GL(d ,{ R})$ vector.  Thus an invariant
combination is
$$({\cal F}^T V)_{\alpha} (V^T \eta V)^{-1}_{\alpha \beta} (V^T {\cal
F})_{\beta}\,\, .\eqn\scn$$
It is straightforward to show that in the $Y = 1$ gauge this reduces to
$${1 \over 2} \big({\cal F}^T  \eta  {\cal F} + {\cal F}^T  M^{-1}
{\cal F}\big) \,\, .\eqn\sco$$
Thus the desired result is
$${\cal L}_3 = - {1 \over 2} ({\cal F}^T V)  (V^T \eta V)^{-1} (V^T
{\cal F}) + {1 \over 4} {\cal F}^T \eta {\cal F} \,\, .\eqn\scp$$

The result found above is not what was expected.  Experience from
supergravity theories leads one to expect that it should be possible to
linearize the $O(d, d)$ symmetry transformations by introducing a
complete adjoint multiplet of scalar fields and gauging the maximal
compact subgroup $O(d) \times O(d)$, so that the $d^2$ scalar fields $X
= G + B$ would parametrize the coset space $O(d, d) /O(d)\times O(d)$.
What we have done above is quite different -- it is not a coset construction,
since the starting multiplet of scalars $V$ does not parametrize the
adjoint representation of any group.  Rather it belongs to vector
representations of both $O(d, d)$ and $GL(d, { R})$, the latter being
gauged.  This raises a question.  Does the usual $G/H$ construction give
an equivalent result or does it give a wrong result?

To construct an $O(d, d)/O(d) \times O(d)$ theory we follow the
procedure used in various supergravity theories.  The way to do this is
to introduce a $2d \times 2d$ matrix $V_{Ai}$ which plays the role of a
``vielbein" for the matrix $M_{ij}$ [\DUFFB], in the sense that
$$M_{ij} = (V^T V)_{ij} = \delta^{AB} V_{Ai} V_{Bj} \,\, .\eqn\scq$$
A matrix that solves this equation is
$$ V = \pmatrix {E^{-1} & -E^{-1} B\cr 0 & E \cr} \eqn\scr$$
where $E$ is a $ d \times d$ vielbein satisfying $E^T E = G$.  It should be
noted that the matrix $V$ belongs to $O(d, d)$, i.e., $V^T \eta V = \eta$.

The obvious guess, then, for an action with global $O(d, d)$ symmetry
and local $O(d) \times O(d)$ symmetry is
$${\cal L} = {1 \over 4}
\eta^{ij} \eta^{AB} (D_{\mu} V)_{Ai} (D^{\mu} V)_{Bj} \,\, ,\eqn\scs$$
with $V$ an arbitrary $O(d, d)$ matrix (not yet of the special form in
eq. \scr) and auxiliary gauge fields for local $O(d) \times O(d)$, which
are incorporated in the covariant derivatives.  The covariant derivative
is a little awkward to formulate in the basis
with the off-diagonal metric $\eta$. Therefore we
make a change of basis that diagonalizes it.  Introducing $\rho^T \eta
\rho = \sigma$, where
$$\rho = {1 \over \sqrt{2}} \pmatrix { 1 & -1 \cr 1 & 1
\cr} \quad {\rm and} \quad
\sigma = \pmatrix {1 & 0 \cr 0 & -1 \cr}\, , \eqn\sct$$
we rotate the
matrix $V$ by defining $W = \rho^T V \rho$.  Since $V$ is an $O(d, d)$
matrix satisfying $V^T \eta V = \eta$, $W$ is an $O(d, d)$ matrix
satisfying $W^T \sigma W = \sigma$.  Now the covariant derivative
takes the form
$$(D_{\mu} W)_{Ai} = \partial_{\mu} W_{Ai} + \omega_{\mu
AB} \sigma^{BC} W_{Ci} \,\, ,\eqn\scu$$
where the auxiliary $O(d) \times O(d)$
gauge fields are given by
$$\omega_{\mu} = \pmatrix {\omega^{(1)}_{\mu}
& 0 \cr 0 & \omega^{(2)}_{\mu} \cr } \,\, .\eqn\scv$$
In this expression
$\omega^{(1)}_{\mu}$ and $\omega^{(2)}_{\mu}$ are independent $O(d)$
gauge fields (antisymmetric).  The Lagrangian now takes the form
$${\cal L} = {1 \over 4} ~ \sigma^{ij} \sigma^{AB} (D_{\mu} W)_{Ai}
(D^{\mu}W)_{Bj}\,\, .\eqn\scw$$

To make contact with ${\cal L}_2$ one varies with respect to the gauge
fields, solves their classical equations, and substitutes back into
${\cal L}$.  This procedure is certainly valid in the present context.
One finds that
$$\omega^{(1)}_{\mu ab} = {1 \over 2}  \eta^{ij} (W_{ai}
\partial_{\mu} W_{bj} - W_{bi} \partial_{\mu} W_{aj})\eqn\scx$$
where $a ,b$ run over the first $d$ values of the indices $A, B$.
$\omega^{(2)}_{\mu ab}$ is given by an analogous formula using the
second $d$ values of the indices.  Substituting back into eq. \scw\ we
find that
$${\cal L} = {1 \over 2} {\rm tr} \big[ ( W \sigma \partial_{\mu} W^T)_{12}
{}~ (W \sigma \partial^{\mu} W^T )_{21} \big] \, .\eqn\scy$$
The notation is that the numerical indices represent
$d \times d$ blocks of the $2d \times 2d$ matrix
$W \sigma \partial_{\mu} W^T$.
At this point $W$ is an arbitrary $O(d, d)$ matrix.  However,
${\cal L}$ still has local $O(d) \times O(d)$ symmetry even though the
gauge fields have been eliminated.  This local symmetry allows us to choose
a gauge in which $W$ takes the form $W = \rho^T V \rho$, with $V$ the
matrix given in eq. \scr.

Now we must compare the result above to
${\cal L}_2 = {1 \over 8} {\rm tr}
(\eta ~ \partial_{\mu} M ~ \eta ~ \partial^{\mu}
M) \, .$  Substituting $M = V^T V = \rho W^T W \rho^T$ and using $\rho^T
\eta \rho = \sigma$ gives
$${\cal L}_2 = {1 \over 8} tr (\sigma ~\partial_{\mu} (W^TW) \sigma
{}~\partial^{\mu} (W^TW))\,\,. \eqn\scz$$
Expanding out the derivatives and using $W^T \sigma W = \sigma$ gives
$${\cal L}_2 = {1 \over 4} {\rm tr} \bigg [ (W \sigma ~\partial_{\mu} W^T)(W
\sigma~\partial^{\mu} W^T) - (W \sigma ~\partial_{\mu} W^T) \sigma (W
\sigma \partial^{\mu} W^T) \sigma \bigg ] \,\,.\eqn \scaa$$
Expanding in $d \times d$ blocks one finds that eqs. \scy\ and \scaa\
are identical, and hence ${\cal L} = {\cal L}_2$,
as desired!  Thus the conventional wisdom that $G$ and $B$ parametrize
an $O(d, d)/O(d) \times O(d)$ coset is correct.  The somewhat
surprising fact is that there is an alternative interpretation
utilizing local $GL (d, { R})$ described in the first part of this
section.

\chapter{\bf Generalization to O(d, d+n) Symmetry}

Previous work in supergravity [\MS] and superstring theory [\Narain]
suggests that if we
add $n$ Abelian $U(1)$ gauge fields to the original $(D + d)$-dimensional
theory, that $O(d, d+n)$ symmetry should result from dimensional
reduction to $D$ dimensions.  In this section we explore whether this is
the case.  The additional term to be added to the action is
$$S_{\hat A} = - {1 \over 4} \int_M dx \int_K dy~ {\sqrt {- \hat g}}~ e^{-
\hat \phi} ~ \hat g^{\hat \mu \hat \rho} \hat g^{\hat \nu \hat \lambda}
 \delta_{IJ} \hat F^I_{\hat \mu \hat \nu} ~ \hat F^J_{\hat \rho \hat
\lambda} \,\, ,\eqn\sda$$
where $\hat F^I_{\hat \mu \hat \nu} = \partial_{\hat \mu} \hat A^I_{\hat
\nu} - \partial_{\hat \nu} \hat A^I_{\hat \mu}$ and the index $I$ takes
the values $I = 1, ~2, \cdots , n$.

The most important point to note is that the original $(D +
d)$-dimensional theory should have $O(n)$ symmetry described by the
formulas of section 2 with $M_{IJ} = \eta_{IJ} = \delta_{IJ}$.  Looking
at the various pieces of the Lagrangian, we see that ${\cal L}_1$ has
the usual form, ${\cal L}_2 = 0, ~ {\rm and}~ {\cal L}_3 ~{\rm gives}~
S_{\hat A}$.  The crucial observation concerns ${\cal L}_4$, which is
built from the square of
$$\hat H_{\hat \mu \hat \nu \hat \rho} = \partial_{\hat \mu}\hat B_{\hat \nu
\hat \rho} - {1 \over 2}\hat A^I_{\hat \mu} \delta_{IJ} \hat F^J_{\nu
\rho} + {\rm (cyc. ~ perms.)}\,\,.\eqn\sdb$$
This contains the Chern--Simons term (for the $U(1)$ gauge fields),
a feature that is clearly crucial for the symmetries we wish to
implement.
\foot{It is remarkable that the necessity of the Chern--Simons terms is
deduced from purely bosonic considerations. This has been argued
previously, in the $\sigma$-model description of strings, based on
anomaly effects arising from gauge fields that couple chirally to the
world sheet [\NEPO]. Such a chiral coupling is not
assumed in the present analysis.}
Once this point is understood, the analysis is a fairly
straightforward generalization of that presented in section 2, though
some of the algebra is more complicated.

The dimensional reduction can now be carried out by the same methods
introduced in section 2.  The reduction of $S_{\hat g}$ is unchanged
from before.  For the vectors we obtain
$$S_{\hat A} = - {1 \over 4} ~\int dx {\sqrt -g}~ e^{- \phi} \bigg\{
F^I_{\mu \nu} F^{I \mu \nu} + 2 F^I_{\mu \alpha} F^{I \mu
\alpha}\bigg\}\,\,,\eqn\sdc$$
where we define
$$\eqalign {
A^{(3)I}_{\mu} &= \hat A_{\mu}^I -a_{\alpha}^I A_{\mu}^{(1)\alpha}\cr
F^{(3)I}_{\mu\nu} &= \partial_{\mu} A^{(3)I}_{\nu} -
\partial_{\nu} A^{(3)I}_{\mu}\cr
a^I_{\alpha} &= \hat A^I_{\alpha} \cr
F^I_{\mu \nu} &= F^{(3) I}_{\mu \nu} + F^{(1) \alpha}_{\mu
\nu} a^I_{\alpha}\cr
F^I_{\mu \alpha}&= \partial_{\mu} a^I_{\alpha}\,\, .\cr}\eqn\sdd$$

The reduction of the various $H$ terms includes additional pieces beyond
those of section 2, because of the presence of the Chern--Simons term.
We find the following:
$$\eqalign {H_{\mu \alpha \beta} &= \partial_{\mu} B_{\alpha \beta} + {1
\over 2} ( a^I_{\alpha} \partial_{\mu} a^I_{\beta} - a^I_{\beta}
\partial_{\mu} a^I_{\alpha})\cr
	H_{\mu \nu \alpha} &= - C_{\alpha \beta} F^{(1) \beta}_{\mu \nu}
	+ F^{(2)}_{\mu \nu \alpha} - a^I_{\alpha} F^{(3) I}_{\mu \nu}\cr
	H_{\mu \nu \rho} &= \partial_{\mu} B_{\nu \rho} - {1 \over 2}
{\cal A}^i_{\mu} \eta_{ij} {\cal F}^j_{\nu \rho} + {\rm cyc.~ perms.}\,\, ,
\cr}\eqn\sde$$
where we have used the definitions
$$A^{(2)}_{\mu \alpha} = \hat B_{\mu \alpha} + B_{\alpha
\beta} A^{(1) \beta}_{\mu} + {1 \over 2} a^I_{\alpha}
A^{(3)I}_{\mu}\eqn\sdf$$
$$C_{\alpha \beta}= {1 \over 2} a^I_{\alpha} a^I_{\beta} +
B_{\alpha\beta}\,\,.\eqn\sdg$$
As usual, $H_{\mu\nu\rho}$ is gauge invariant for $\delta {\cal
A}^i_{\mu} = \partial_{\mu} \Lambda^i$ and $\delta
B_{\mu\nu}={1\over 2}\Lambda^i \eta_{ij}{\cal F}^j_{\mu\nu}$.
In matrix notation we write $C = {1 \over 2} a^T a + B.$
We have introduced a $(2d + n)$-component vectors ${\cal A}_{\mu}^i
= (A_{\mu}^{(1) \alpha} ,
{}~ A_{\mu\alpha}^{(2)} , ~ A_{\mu}^{(3)I})$ and ${\cal F}^i_{\mu\nu}=
\partial_{\mu}{\cal A}^i_{\nu} -
\partial_{\nu}{\cal A}^i_{\mu}$ and the $O(d, d + n)$ metric $\eta$,
which, when written in blocks, takes the form
$$ \eta = \pmatrix {0 & 1 & 0 \cr 1 & 0 & 0 \cr 0 & 0 & 1 \cr} \,\,.
\eqn\sdh$$
With these definitions, $H_{\mu \nu \rho}$ has manifest $O(d, d + n)$
symmetry.

Next, we look at all terms that are quadratic in field strengths $F$.
The contributions to
$${\cal L}_3 = - {1 \over 4} ~ {\cal F}^i_{\mu \nu} (M^{- 1})_{ij} ~
{\cal F}^{j\mu \nu}
\eqn\sdi$$
come from $S_{\hat g}$ (as before), from ${1 \over 4} F^I_{\mu \nu} F^{I
\mu \nu}$, and from ${1 \over 4} H_{\mu \nu \alpha}H^{\mu \nu \alpha}$.
{}From these we
read off the result
$$M^{-1 } = \pmatrix {G + C^T G^{-1} C + a^Ta & -C^T G^{-1} & C^T G^{-1}
a^T + a^T \cr
	-G^{-1} C & G^{-1} & - G^{-1} a^T \cr
		a G^{-1} C + a & -a G^{-1 } & 1 + a G^{-1} a^T\cr }\eqn\sdj$$
To check whether this is an $O (d , d+n)$ matrix we form
$$\eta M^{-1} \eta = \pmatrix {G^{-1} & - G^{-1} C & - G^{-1} a^T \cr
	- C^T G^{-1} & G + C^T G^{-1} C + a^T a & C^T G^{-1} a^T + a^T \cr
	- a G^{-1} & a G^{-1} C + a & 1 + a G^{-1} a^T \,\, .\cr} \eqn\sdk$$
Multiplying these, we find that $M^{-1} \eta M^{-1} \eta = 1$.  Hence
$M^{-1}$ and $M$ are symmetric $O(d, d+n)$ matrices, as expected.

Motivated by the results of section 3, we next seek a matrix $V$
belonging to $O(d, d+n)$ such that $V^TV = \eta M^{-1} \eta = M$.  It is
very easy at this point to discover that a suitable choice is
$$V = \pmatrix {E^{-1} & - E^{-1} C & - E^{-1} a^T \cr
	0 & E & 0 \cr
	0 & a & 1 \cr}\,\, ,\eqn\sdl$$
which is remarkably simple.

The last remaining check of $O(d, d+n)$ symmetry is to verify that we
recover ${\cal L}_2 = {1 \over 8} {\rm tr} (\partial_{\mu} M^{-1}
\partial^{\mu} M)$, with the matrix $M$ given above.  Relevant
contributions come from $S_{\hat g} , ~ - {1 \over 2} (F^I_{\mu
\alpha})^2$, and $- {1 \over 4} (H_{\mu \alpha \beta})^2$.  The
calculation is a bit tedious, but the desired result is obtained.
Clearly this term can also be understood using an $O(d, d+n)/O(d) \times
O(d + n)$ analysis generalizing that presented in section 3.  In this
case ${\cal L}_2$ would be obtained, as before, if one uses the local
$O(d) \times O(d + n)$ symmetry to bring an arbitrary $O(d, d+n)$ matrix
$V$ to the form given above.

It is natural to inquire what happens if the $(D + d)$-dimensional
theory contains a non-Abelian Yang--Mills group.  After all, the
heterotic string in 10 dimensions can have $O(32)$ or $E_8 \times E_8$.
In general, compactification with nontrivial moduli breaks these
symmetries.  The only thing that is easy to do, and which makes contact
with Narain's analysis [\Narain],
is to set to zero all the gauge fields except
those belonging to a Cartan subalgebra -- $[U(1)]^{16}$ in the case of the
heterotic string.  Then the problem reduces to the Abelian theory, and
the analysis of this section becomes applicable.  In this way one
obtains the noncompact group $O(d, d+16)$ considered by Narain.

\chapter{\bf O(d,d) Symmetric World Sheet Equations}

In this section, we discuss how $O(d, d)$ symmetry appears in the
$\sigma$-model description of string theory. The generalization to the
$O(d,d+n)$ case will not be presented in detail here, but the result
will be stated at the end of the section.  First, we review the
description of a string in the presence of constant background fields
when $d$ coordinates are compactified on a torus,
following refs. [\NSW,\SW,\JHS].
Then we consider the extension to spacetime-dependent background fields,
generalizing previous studies in refs.
[\DUFFA,\CFG,\MO,\DUFFB,\JM,\KKK]. Only closed string theories are
considered here. For a recent discussion of toroidal compactification of
open string theories see [\BIANCHI].

To be specific, let us consider the two-dimensional $\sigma$-model
description of a bosonic string in a space with
$d$ compactified coordinates $Y^{\alpha}(\sigma,\tau)$. The portion of
the action containing these coordinates is
$$S_K={1\over 2} \int d^2 \sigma ~ \big[ G_{\alpha\beta} \eta^{ab}\partial_{a}
Y^{\alpha}
\partial_{b} Y^{\beta} + \epsilon^{ab} B_{\alpha\beta} \partial_{a}
Y^{\alpha}
\partial_{b} Y^{\beta} \big]\, , \eqn\seaa$$
where $G_{\alpha\beta}$ and $B_{\alpha\beta}$ are constants.
The coordinates are taken to satisfy the periodicity conditions
$Y^{\alpha} \simeq Y^{\alpha} + 2 \pi$.
\foot{We apologize for switching conventions from section 2, where the
$y^{\alpha}$'s were taken to have unit periods.}
For closed strings it is necessary that
$$Y^{\alpha} (2 \pi , \tau) = Y^{\alpha} ( 0, \tau) + 2 \pi m^{\alpha}\,
,\eqn\sea$$
where the integers $m^{\alpha}$ are called winding numbers. It follows from
the single-valuedness of the wave function on the torus that the zero modes
of the canonical momentum, $P_{\alpha} = G_{\alpha\beta}
\partial_{\tau} Y^{\beta} + B_{\alpha\beta}
\partial_{\sigma} Y^{\beta}$, are also integers $n_{\alpha}$. Therefore
the zero modes of $Y^{\alpha}$ are given by
$$Y^{\alpha}_0 = y^{\alpha} + m^{\alpha} \sigma + G^{\alpha\beta}
(n_{\beta} - B_{\beta\gamma} n^{\gamma}) \tau \, , \eqn\seab$$
where $G^{\alpha\beta}$ is the inverse of $G_{\alpha\beta}$ as before.
The Hamiltonian is given by
$${\cal H} = {1\over 2} G_{\alpha\beta} ( \dot Y^{\alpha} \dot
Y^{\beta} + Y'^{\alpha} Y'^{\beta} )\, , \eqn\seac$$
where $\dot Y^{\alpha}$ and $Y'^{\beta}$ are
derivatives with respect to $\tau$ and $\sigma$, respectively.

Since $Y^{\alpha} (\sigma , \tau)$ satisfies the free wave equation, we can
decompose it as the sum of left- and right-moving pieces.
The zero mode of $P^{\alpha}=G^{\alpha\beta}P_{\beta}$
is given by $p_L^{\alpha}+p_R^{\alpha}$ where
$$p^{\alpha}_L = {1\over 2}
[ m^{\alpha} + G^{\alpha\beta} (n_{\beta} - B_{\beta\gamma} m^{\gamma}) ]
\quad {\rm and} \quad p^{\alpha}_R = {1\over 2}
[ - m^{\alpha} + G^{\alpha\beta} (n_{\beta} - B_{\beta\gamma}
m^{\gamma} ) ]\eqn\seb$$
The mass-squared operator, which corresponds to the zero mode of ${\cal H}$,
is given (aside from a constant) by
$$(mass)^2 = G_{\alpha\beta} \big( p^{\alpha}_L p^{\beta}_L + p^{\alpha}_R
p^{\beta}_R \big) +
\sum^{\infty}_{m=1}\sum_{i=1}^d (\alpha^i_{- m} \alpha^{i}_m + \tilde
\alpha^{i}_{- m}
\tilde\alpha^{i}_m)  \, , \eqn\sed$$
As usual, $\{\alpha_m\}$
and $\{ \tilde \alpha_m\}$ denote oscillators associated with
right- and left-moving coordinates,
respectively. Substituting the expressions for $p_L$ and $p_R$,
the mass squared can be rewritten as
$$(mass)^2 = {1\over 2} G_{\alpha\beta}
m^{\alpha} m^{\beta} + {1\over 2} G^{\alpha\beta} (n_{\alpha} -
B_{\alpha\gamma} m^{\gamma})(n_{\beta} - B_{\beta\delta} m^{\delta})
+\sum (\alpha^i_{- m} \alpha^i_m + \tilde \alpha^i_{- m}
\tilde \alpha^i_m) \, . \eqn\sec$$
It is significant that the zero mode portion of eq. \sec\ can be
expressed in the form
$$(M_0)^2 = {1\over 2}
(m \ \ n)  M^{-1} \pmatrix {m\cr n\cr},\eqn\sex$$
where $M$ is the $2d\times2d$ matrix introduced in section 2, which we
display once again:
$$M = \pmatrix {G^{-1} & -G^{-1} B\cr
BG^{-1} & G - BG^{-1} B\cr}\eqn\sbq$$
In order to satisfy
$\sigma$-translation symmetry, the contributions of left- and
right-moving sectors to the mass squared
must agree ($L_{0}=\tilde L_{0}$ in the
usual notation). The zero mode contribution to
their difference is
$$G_{\alpha\beta} (p^{\alpha}_L p^{\beta}_L - p^{\alpha}_R p^{\beta}_R )
= m^{\alpha} n_{\alpha} \, . \eqn\sey$$
Since this is an integer, it always can be compensated by oscillator
contributions, which are also integers.

Equation \sey\ is invariant under interchange of
the winding numbers $m^{\alpha}$ and the discrete momenta $n_{\alpha}$.
Indeed, the
entire spectrum remains invariant if we interchange
$m^{\alpha} \leftrightarrow
n_{\alpha}$ and simultaneously let [\SW]
$$(G - B G^{-1} B) \leftrightarrow G^{- 1} \quad {\rm
and} \quad B G^{- 1} \leftrightarrow - G^{- 1} B \, . \eqn\see$$
These interchanges precisely correspond to inverting the $2d\times2d$
matrix $M$. This is the spacetime duality
transformation generalizing the well-known duality $R\leftrightarrow
\alpha'/R$ in the $d=1$ case [\GSB,\KY,\NSSW,\RV].  The general duality
symmetry
implies that the $2d$-dimensional Lorentzian lattice
spanned by the vectors ${\sqrt 2}
(p^{\alpha}_L , \, p^{\alpha}_R)$ with inner product
$${\sqrt 2}~ (p_L , \, p_R) \cdot
{\sqrt 2}~ (p'_L , \, p'_R) \equiv 2G_{\alpha\beta} (p^{\alpha}_L p'^{\beta}_L
-
p^{\alpha}_R p'^{\beta}_R) = (m^{\alpha} n'_{\alpha} + m'^{\alpha}
n_{\alpha})\, , \eqn\sead$$
is even and self-dual [\Narain].

The moduli space parametrized by $G_{\alpha\beta}$ and $B_{\alpha\beta}$
is locally the coset
$O(d, d)/O(d) \times O(d)$ [\NSW], just as we found in section 3.
The global geometry requires also modding out the group of discrete symmetries
generated by $B_{\alpha\beta} \rightarrow B_{\alpha\beta} +
N_{\alpha\beta}$ and $G + B \rightarrow
(G + B)^{-1}$.  These symmetries generate the $O(d,d,Z)$ subgroup of
$O(d,d)$. An $O(d,d,Z)$ transformation is given by a $2d\times2d$ matrix $A$
having integral entries and satisfying $A^T \eta A = \eta$, where $\eta$
consists of off-diagonal unit matrices as before. Under an $O(d,d,Z)$
transformation
$$\pmatrix {m \cr n} \rightarrow \pmatrix {m' \cr n'} =
A \pmatrix {m \cr n \cr}
\quad {\rm and} \quad M \rightarrow AMA^T\ .\eqn\sef$$
It is evident that
$$m\cdot n = {1\over 2}(m \ \ n) \eta \pmatrix {m
\cr n \cr}\, , \eqn\seg$$
which appears in eq. \sey,
and $M_0^2$ in eq. \sex\ are preserved under these transformations.
The crucial fact, already evident from the spectrum, is that toroidally
compactified string theory certainly does not share the full $O(d,d)$
symmetry of the low energy effective theory. It is at most invariant
under the discrete $O(d,d,Z)$ subgroup. However, as emphasized by Sen
[\HS], if the $Y$ coordinates are {\it not} compactified, but still flat,
so that $K=R^d$, there
is a continuous $O(d) \times O(d)$ symmetry (the compact part of
$O(d,d)$) corresponding to independent rotations of $Y_L$ and $Y_R$. The
diagonal subgroup describes ordinary rotations of $K$.

Now we turn our attention to the case when the
$(D+d)$-dimensional massless background fields
$\hat g_{\hat \mu \hat \nu}$ and $\hat B_{\hat \mu \hat \nu}$
depend on $D$ coordinates. The $D+d$ string
coordinates $X^{\hat \mu}$ decompose into two sets $\{X^{\mu}\}$ and
$\{Y^{\alpha}\}$ where $\mu = 0, 1, \dots, D - 1$ and $\alpha = 1, 2,
\dots, d$.
The world sheet action is
$$	S = {1\over 2} \int d^2\sigma (\hat g_{\hat \mu \hat\nu}
\eta^{ab} + \hat B_{\hat\mu \hat\nu} \epsilon^{ab})\partial_a
X^{\hat\mu} \partial_b X^{\hat\nu} \,\, .\eqn\seea $$
Varying this with respect to $X^{\hat\mu} (\sigma, \tau)$ gives the
classical equation of motion for the string
$$\eqalign{{\delta S\over\delta X^{\hat\mu}} =\ & -\hat\Gamma_{\hat\mu \hat\nu
\hat\rho} \partial^a X^{\hat\nu} \partial_a X^{\hat\rho} - \hat
g_{\hat\mu \hat\nu} \partial^a \partial_a X^{\hat\nu}\cr
\ & + {1\over 2} \epsilon^{ab} (\partial_{\hat\mu} \hat B_{\hat\nu
\hat\rho} + \partial_{\hat\nu} \hat B_{\hat\rho \hat\mu} +
\partial_{\hat\rho} \hat B_{\hat\mu \hat\nu}) \partial_a X^{\hat\nu}
\partial_b X^{\hat \rho} = 0\ ,\cr}
\eqn\seeb$$
where
$$	\hat\Gamma_{\hat\mu \hat\nu \hat\rho} = {1\over 2}
( \partial_{\hat\nu} \hat
g_{\hat\mu \hat\rho} + \partial_{\hat\rho} \hat
g_{\hat\mu \hat\nu} - \partial_{\hat\mu} \hat g_{\hat\nu \hat\rho})
 \,\, .\eqn\seec$$
To analyze these equations it is convenient to consider $X^\mu$ and
$Y^\alpha$ separately.  Since the $Y^\alpha$ equation is somewhat
simpler we begin with that.  Indeed for that case, let us back up and
focus on those terms in $S$ that are $Y$ dependent.  These are
$$S_Y = \int d^2 \sigma \bigg\{ {1\over 2} \big(\eta^{ab}G_{\alpha \beta}(X)
\partial_a Y^{\alpha}
\partial_b Y^{\beta}  + \epsilon^{ab}B_{\alpha \beta}(X) \partial_a
Y^{\alpha} \partial_b Y^{\beta} \big) + {\Gamma}^a_{\alpha}(X)
\partial_a Y^{\alpha}\bigg\}\, ,\eqn\sei$$
where
$$\eqalign{{\Gamma}^a_{\alpha} \equiv\ & \eta^{ab} \hat g_{\mu
\alpha} \partial_b X^{\mu} - \epsilon^{ab}
\hat B_{\mu \alpha} \partial_b X^{\mu} \cr
=\ & \eta^{ab} G_{\alpha\beta}
A^{(1)\beta}_{\mu}\partial_b X^{\mu} - \epsilon^{ab}
\big( A^{(2)}_{\mu\alpha} - B_{\alpha\beta}
A^{(1)\beta}_{\mu}\big)\partial_b X^{\mu} \cr} \eqn\sej$$
encodes information about the gauge fields $A^{(1)\alpha}_{\mu}$ and
$A^{(2)}_{\mu\alpha}$. This action
generalizes eq. \seaa, both by including background vector fields and by
allowing $X$ dependence for all the background fields.

The goal now is
to study the resulting $Y$ equations of motion, to recast them into
a form with manifest $O(d,d)$ symmetry, and to understand why the
symmetry breaks to $O(d,d,Z)$. The $O(d,d)$ symmetry cannot be explicitly
realized on the action. Rather, it is necessary to combine the equations of
motion for $Y$ with those for dual coordinates $\tilde Y$,
in order to make the symmetry manifest [\CFG,\DUFFB].
\foot{Previous studies of four-dimensional examples illustrate
that, when duality transformations are required,
the equations of motion can be made manifestly invariant
under the ``hidden" symmetries even though the action cannot be [\GZ].
Actually, hidden symmetries sometimes can be made manifest in the action
it one is willing to give up some other symmetry. For example, in ref.
[\AT], a duality invariant action that does not have manifest world
sheet Lorentz invariance is formulated.}
In the absence of nontrivial backgrounds, $Y$ and $\tilde Y$ would
correspond to the sum and difference of left-moving and right-moving
components. In more general settings, the interpretation is not quite so
simple. It has been suggested on occasion [\EDWARD,\DUFFB,\AT] that this
doubling of coordinates has some deep significance. However one feels
about that, the mathematics is indisputable.

Since the backgrounds are independent of $Y^{\alpha}$, the
Euler--Lagrange equations take the form
$$\partial_a \bigg({\delta S \over \delta \partial_a
Y^{\alpha}} \bigg) = 0 \, . \eqn\sek$$
Therefore, locally, we can write
$${\delta S \over \delta \partial_a Y^{\alpha}} = \eta^{ab}
\partial_b Y^{\beta} G_{\alpha \beta} + \epsilon^{ab} \partial_b Y^{\beta}
B_{\alpha \beta} + {\Gamma}^a_{\alpha} = \epsilon^{ab}
\partial_b \tilde Y_{\alpha}  \,\,, \eqn\sel$$
where $\tilde Y_{\alpha}$ are the dual coordinates. They clearly have
the same periodicities as the $Y^{\alpha}$. Introducing auxiliary fields
$U^{\alpha}_a$, let us now define a dual action
$$\tilde S = \int d^2\sigma \bigg\{{1 \over 2} \big( \eta^{ab} U^{\alpha}_a
U^{\beta}_b G_{\alpha \beta} + \epsilon^{ab} U^{\alpha}_a
U^{\beta}_b B_{\alpha \beta}\big) + \epsilon^{ab} \partial_a \tilde Y_{\alpha}
U^{\alpha}_b + {\Gamma}^a_{\alpha} U^{\alpha}_a\bigg\} \, . \eqn\sem$$
Varying this action with respect to $\tilde Y_{\alpha}$
gives $\partial_a(\epsilon^{ab} U^{\alpha}_b)=0$, while the $U^{\alpha}_a$
equation of motion
$$\eta^{ab} U^{\beta}_b G_{\alpha \beta} + \epsilon^{ab}
U^{\beta}_b B_{\alpha \beta} - \epsilon^{ab} \partial_b \tilde Y_{\alpha} +
{\Gamma}^a_{\alpha} = 0 \eqn\sen$$
agrees with eq. \sel\ when one identifies $U^{\alpha}_a$ with
$\partial_a Y^{\alpha}$.
This can be used to solve for $U^{\alpha}_a$ in terms of $\partial_a
\tilde Y_{\alpha}$ and ${\Gamma}^a_{\alpha}$. The result is
$$U^{\alpha}_a = \big(\epsilon_a{}^b
{\cal G}^{\alpha \beta} +\delta^b_a {\cal B}^{\alpha \beta}\big)
\big( \partial_b\tilde Y_{\beta} - \epsilon_{bc} \Gamma^c_{\beta}\big)
 \, , \eqn\seo$$
where we have introduced
$${\cal G} = (G - BG^{-1}B)^{-1} \eqn\secalg$$
and
$${\cal B} = - G^{-1}B (G - BG^{-1}B)^{-1}\, . \eqn\secalb$$
Note that $(G+B)({\cal G}+{\cal B})=1$, so that ${\cal G}$ and ${\cal
B}$ are the symmetric and antisymmetric parts of $(G+B)^{-1}$,
respectively.

Substituting for $U^{\alpha}_a$, the dual action \sem\ takes the form
$$\eqalign {\tilde S =& \int d^2\sigma \bigg\{{1\over 2}\big( \eta^{ab}
\partial_a \tilde Y_{\alpha} \partial_b \tilde Y_{\beta} {\cal G}^{\alpha
\beta} + \epsilon^{ab} \partial_a \tilde Y_{\alpha}
\partial_b \tilde Y_{\beta} {\cal B}^{\alpha \beta}\big)
- \epsilon^a{}_b \partial_a
\tilde Y_{\alpha} {\Gamma}^b_{\beta} {\cal G}^{\alpha \beta}\cr &
- \partial_a \tilde Y_{\alpha} {\Gamma}^a_{\beta} {\cal B}^{\alpha \beta}
- {1 \over 2}\big( \eta_{ab}
{\Gamma}^a_{\alpha} {\Gamma}^b_{\beta} {\cal G}^{\alpha \beta} +
\epsilon_{ab} {\Gamma}^a_{\alpha} {\Gamma}^b_{\beta} {\cal B}^{\alpha
\beta}\big)\bigg\}\, .\cr} \eqn\sep$$
Since ${\cal G}^{\alpha \beta}$ and ${\cal
B}^{\alpha \beta}$ are determined in terms of $G_{\alpha \beta}$ and
$B_{\alpha \beta}$, they depend only on $X^{\mu}$, as
does ${\Gamma}^a_{\alpha}$.  As before, the equation of motion derived
from $\tilde S$ is $\partial_a \bigg( {\delta \tilde S \over \delta
\partial_a \tilde Y_{\alpha}} \bigg) = 0$.
The two Lagrangians $S$ and $\tilde S$ give a pair of equivalent
equations of motion (at least locally), which are obtained by applying
$\partial_a$ to eq. \sel\ and
$$\epsilon^{ab} \partial_b Y^{\alpha} = {\delta \tilde S \over
\delta \partial_a \tilde Y_{\alpha}} = \eta^{ab}
\partial_b \tilde Y_{\beta} {\cal G}^{\alpha \beta} + \epsilon^{ab}
\partial_b \tilde Y_{\beta} {\cal B}^{\alpha \beta} - \epsilon^a{}_b {\cal
G}^{\alpha \beta} {\Gamma}^b_{\beta} - {\cal B}^{\alpha \beta}
{\Gamma}^a_{\beta}\, . \eqn\ser$$
In order to express equations \sel\ and \ser\ in an $O(d, d)$
covariant form, let us multiply them by $G^{-1}$ and ${\cal G}^{-1}$,
respectively, as well as by $\epsilon^{ab}$, as follows:
$$G^{\alpha \beta} \partial_a \tilde Y_{\beta} -
( G^{-1} B)^{\alpha}{}_{\beta} \partial_a Y^{\beta} = \epsilon_a{}^b
\partial_b Y^{\alpha} + \epsilon_{ab} G^{\alpha \beta}
{\Gamma}^b_{\beta}\,\, .\eqn\ses$$
$$({\cal G}^{-1})_{\alpha \beta} \partial_a Y^{\beta} - ({\cal G}^{-1}
{\cal B})_{\alpha}{}^{\beta} \partial_a \tilde Y_{\beta} =
\epsilon_a{}^b \partial_b \tilde Y_{\alpha} - \eta_{ab}
{\Gamma}^b_{\alpha} - \epsilon_{ab} ({\cal G}^{-1}
{\cal B})_{\alpha}{}^{\beta}\Gamma^b_{\beta} \, . \eqn\set$$
If we define an enlarged manifold combining the coordinates $Y^{\alpha}$
and $\tilde Y_{\alpha}$ such that $\{ Z^i\} = \{Y^{\alpha} , \, \tilde
Y_{\alpha} \}, \, i = 1, 2, \dots, 2d$, then eqs. \ses\ and \set\
can be combined as the single equation
$${ M} \eta  \partial_a Z = \epsilon_a{}^b
\partial_b Z + {M} \eta \Sigma_a \, . \eqn\seu$$
Here $\Sigma_a$ is an $O(d, d)$ vector (for each value of $a$)
given by the column vector
$$\Sigma^i_a = \pmatrix {- \eta_{ab} G^{\alpha \beta} {\Gamma}^b_{\beta} &\cr
\epsilon_{ab} {\Gamma}^b_{\alpha} - \eta_{ab}
B_{\alpha \gamma} G^{\gamma \beta} {\Gamma}^b_{\beta}}\, . \eqn\sev$$

Substituting eq. \sej\ into eq. \sev\ gives
$$	\Sigma_a^i = - \partial_a X^\mu {\cal A}_\mu^i + \epsilon_a{}^b
\partial_b X^\mu (M\eta {\cal A}_\mu)^i\,\, , \eqn\seed$$
where ${\cal A}_\mu^i$ is comprised of $A_\mu^{(1)\alpha}$ and
$A^{(2)}_{\mu\alpha}$, as in section 2.  Inserting this into eq. \seu\
then gives the first-order equation
$$	M\eta (\partial_a Z + {\cal A}_\mu \partial_a X^\mu) =
\epsilon_a{}^b (\partial_b Z + {\cal A}_\mu \partial_b X^\mu)\,\, .
\eqn\seee$$
(This equation appears in ref. [\DUFFB] for the special case ${\cal
A}_{\mu}=0$.) One can eliminate $\tilde Y$, of course, obtaining a
second-order equation for $Y$, but then the noncompact symmetry is no
longer evident. This is reminiscent of the issue of making Lorentz
invariance manifest for the Dirac equation. Unlike that case, there is
no obvious action principle that gives the desired
first-order equation for the $Z$ coordinates.
In terms of light-cone components on the world sheet, \seee\ is equivalent
to the pair of equations
$$\eqalign{(1 + M\eta) (\partial_+ Z + {\cal A}_\mu \partial_+ X^\mu)& =
0\cr (1 - M\eta) (\partial_- Z + {\cal A}_\mu \partial_- X^\mu)& = 0
\ .\cr}\eqn\seef$$
These equations have nontrivial solutions, since $(M\eta)^2 = 1$.
Furthermore, they have
manifest $O(d,d)$ invariance provided the transformation
rules $M \rightarrow \Omega M \Omega^T$ and ${\cal A}_\mu \rightarrow
\Omega {\cal A}_\mu$, obtained in section 2, are supplemented
with $Z \rightarrow \Omega Z$.

Using the identity $\eta V\eta V^T = 1$, and recalling that $M = V^T V$,
we can rewrite \seef\ in the form
$$	(\eta \pm 1) V\eta (\partial_\pm Z + {\cal A}_\mu \partial_\pm
X^\mu) = 0 \,\, .\eqn\seeg$$
Written this way, it is clear that the plus and minus cases each consist
of $d$ linearly independent equations.  Defining
$$(D_a Z)^i = \partial_a
Z^i + {\cal A}_\mu^i \partial_a X^\mu \ ,\eqn\seeh$$
the component equations for $Y$ and $\tilde Y$ are
$$\eqalign{(G - B) D_+ Y + D_+ \tilde{Y}& = 0\cr
	(G + B) D_- Y - D_- \tilde{Y}& = 0 \,\, ,\cr}\eqn\seei$$
which is quite a bit simpler than the second order equation for $Y$
that we started from.  Even though these equations have continuous
$O(d,d)$ invariance, the symmetry is broken to the discrete subgroup
$O(d,d,Z)$ by the boundary conditions $Y^\alpha \simeq Y^\alpha + 2\pi$
and $\tilde{Y}_\alpha \simeq \tilde{Y}_\alpha + 2\pi$. The fundamental
point is that all geometries related by $O(d,d,Z)$ transformations
correspond to the same conformal field theory and are physically
equivalent. The moduli space of conformally inequivalent (and hence
physically inequivalent) classical
solutions is given by the coset space $O(d,d)/O(d)\times
O(d)\times O(d,d,Z)$ and is parametrized locally by the scalar fields
$G_{\alpha\beta}$ and $B_{\alpha\beta}$.

The combination $D_aZ^i=\partial_a Z^i + {\cal A}_\mu^i \partial_a
X^\mu$, which appears above, can be given a covariant
interpretation under gauge transformations.  For this purpose it is
necessary to redefine the internal coordinates $Y^\alpha$ and
$\tilde{Y}_a$ in an $X^\mu$ dependent way.  Namely, a gauge
transformation $\delta {\cal A}_\mu^i (X) = \partial_\mu \Lambda^i (X)$ should
be accompanied by $\delta Z^i = - \Lambda^i (X)$.  Despite superficial
appearances, this does not allow the internal coordinates to be
eliminated as part of a gauge choice.  In particular, the winding
numbers $m^\alpha$ and discrete moment $n_\alpha$ are encoded in
$Y^\alpha (2\pi, \tau) = Y^\alpha(0,\tau) + 2\pi m^\alpha$ and
$\tilde{Y}_\alpha (2\pi, \tau) = \tilde{Y}_\alpha (0,\tau) + 2\pi
n_\alpha$.  They cannot be changed by a gauge transformation, since
$X^\mu (2\pi,\tau) = X^\mu (0,\tau)$.

Let us turn now to the $X^\mu$ equation of motion.  This requires
considering eq. \seeb\ for the case of $\hat\mu = \mu$ and substituting
the various definitions given in section 2.  After a certain amount of
algebra one finds, separating different powers of $Y$, that
$$	{\delta S\over\delta X^\mu} = E_\mu^2 + E_\mu^1 + E_\mu^0 \,\,
,\eqn\seej$$
where
$$E_\mu^2 = \partial_\mu (G + B)_{\alpha\beta} \partial_+
Y^\alpha \partial_- Y^\beta \eqn\seek$$
$$\eqalign{E_\mu^1 = &- (A_\mu^{(1)} G)_\alpha
\partial^a \partial_a Y^\alpha +
\epsilon^{ab} \partial_a X^\nu F_{\mu\nu\alpha}^{(2)} \partial_b Y^\alpha\cr
&+ (\partial_\mu [A_\nu^{(1)} (G + B)]_\alpha - (\mu\nu)) \partial_+
X^\nu \partial_- Y^\alpha\cr
&+ (\partial_\mu [A_\nu^{(1)} (G - B)]_\alpha - (\mu\nu)) \partial_-
X^\nu \partial_+ Y^\alpha\cr}\eqn\seel$$
$$\eqalign {E_\mu^0 =\ & - \hat\Gamma_{\mu\nu\rho}
\partial^a X^\nu \partial_a X^\rho -
\hat g_{\mu\nu} \partial^a \partial_a X^\nu\cr
&+ {1\over 2} \epsilon^{ab} (\partial_\mu \hat B_{\nu\rho} +
\partial_\nu \hat B_{\rho\mu} + \partial_\rho \hat B_{\mu\nu})
\partial_a X^\nu \partial_b X^\rho\,\, .\cr}\eqn\seem$$
In the expression for $E_\mu^0$ one must still substitute (see section 2)
$$\hat g_{\mu\nu} = g_{\mu\nu} + A_\mu^{(1)} GA_\nu^{(1)}\eqn\seen$$
and
$$	\hat B_{\mu\nu} = B_{\mu\nu} - {1\over 2} A_\mu^{(1)}
A_\nu^{(2)} + {1\over 2} A_\nu^{(1)} A_\mu^{(2)} + A_\mu^{(1)} B
A_\nu^{(1)} \,\, .\eqn\seeo$$

Now we must try to reexpress all this in an $O(d,d)$ invariant form.  As
a first step consider the manifestly $O(d,d)$ invariant expression
$$ F_\mu^2 = {1\over 2} D_+ Z^i \big(\partial_\mu M^{-1}\big)_{ij} D_-
Z^j \,\, .\eqn\seep$$
Inserting the matrix $M^{-1}$ and expanding out the terms, one can show
(by using the equations of motion \seei) that
$$	F_\mu^2 = D_+ Y ~ \partial_\mu (G + B)~ D_- Y \,\, .\eqn\seeq$$
Therefore, comparing with eq. \seek, we see that
$F_\mu^2$ is an $O(d,d)$ invariant term containing
$E_\mu^2$.  To proceed we must compensate for the terms linear in $Y$ and
independent of $Y$ in $F_\mu^2$ as additions $E_\mu^{1'}$ and
$E_\mu^{0'}$ to $E_\mu^1$ and $E_\mu^0$.  The difference of eqs. \seeq\
and \seek\ gives
$$ \eqalign{E_\mu^{1'} = &- \partial_+ X^\nu A_\nu^{(1)}
\partial_\mu (G + B) ~ \partial_- Y\cr
&- \partial_- X^\nu A_\nu^{(1)} \partial_\mu (G - B)~ \partial_+
Y\cr}\eqn\seer$$
$$ E_\mu^{0'} = - \partial_+ X^\nu A_\nu^{(1)} \partial_\mu (G + B)~
A_\rho^{(1)} \partial_- X^\rho \,\, .\eqn\sees$$

Next, we need to find $O(d,d)$ invariant terms that contain $E_\mu^1 +
E_\mu^{1'}$.  Making repeated use of eq. \seei,
we find that these terms are completely contained in the manifestly
invariant term
$$	F_\mu^1 = \epsilon^{ab} \partial_a X^\nu {\cal F}_{\mu\nu}~ \eta
D_b Z \,\, . \eqn\seet$$
Compensating for the additional $Y$-independent terms that have been
introduced gives
$$\eqalign{E_\mu^{0''} = &- \partial_+ X^\nu \partial_- X^\rho
[A_\mu^{(1)} F_{\nu\rho}^{(2)} + A_\rho^{(1)} F_{\mu\nu}^{(2)} +
A_\nu^{(1)} F_{\rho\mu}^{(2)}\cr
&+ A_\rho^{(1)} (G - B) F_{\mu\nu}^{(1)} + A_\nu^{(1)} (G + B)
F_{\mu\rho}^{(1)}]\cr
&+ \partial_+ [A_\nu^{(1)} \partial_- X^\nu (G - B)] A_\mu^{(1)}
+ \partial_- [A_\nu^{(1)} \partial_+ X^\nu (G + B)]A_\mu^{(1)}\,\,
.\cr}\eqn\seeu$$
To complete this part of the story $E_\mu^0 + E_\mu^{0'} + E_\mu^{0''}$
must still be recast in $O(d,d)$ invariant form.  Remarkably, there is a
great deal of cancellation and one ends up with
$$	F_\mu^0 = - \Gamma_{\mu\nu\rho} \partial^a X^\nu \partial_a X^\rho
- g_{\mu\nu} \partial^a \partial_a X^\nu
+ {1\over 2} \epsilon^{ab} H_{\mu\nu\rho} \partial_a X^\nu
\partial_b X^\rho \,\, ,\eqn\seev$$
with $H_{\mu\nu\rho}$ as defined in section 2.

To summarize, we have found that the $X^\mu$ equation of motion can be
written in the manifestly $O(d,d)$ invariant form
$$\eqalign{ &{1\over 2} D_+ Z \big(\partial_\mu M^{-1}\big)
D_- Z + \epsilon^{ab}
\partial_a X^\nu {\cal F}_{\mu\nu}~ \eta D_b Z \cr
& - \Gamma_{\mu\nu\rho} \partial^a X^\nu \partial_a X^\rho -
g_{\mu\nu} \partial^a \partial_a X^\nu
 + {1\over 2} \epsilon^{ab} H_{\mu\nu\rho} \partial_a X^\nu
\partial_b X^\rho = 0 \,\, .\cr}\eqn\seew$$
Together with eq. \seeg\ or eq. \seei\ this gives the classical dynamics
of strings moving in an arbitrary $X$-dependent background. The
equations are remarkably simple considering all the information they
encode. Clearly, $O(d,d)$ is a useful guide for making them intelligible.

It should come as no surprise to the reader to learn that eqs. \seeg\
and \seew\ continue to hold for the $O(d,d + n)$ generalization,
provided that $M,\eta$, and ${\cal A}_\mu^i$ are defined as in section
4.  Also, $Z^i$ now becomes a $(2d + n)$-component vector made by
combining $Y^\alpha, \tilde{Y}_\alpha$, and $Y^I$, where $Y^I$ are $n$
additional internal coordinates.  It is natural to require that
$$	\partial_- Y^I + A_\mu^{(3)I} \partial_- X^\mu = 0 \,\, ,\eqn\seex$$
as a ``gauge invariant'' generalization of what we know to be true for
the heterotic string with vanishing $A_\mu^{(3)I}$ background fields,
{\it viz.} that the $Y^I$ are left-moving. (The second term in eq.
\seex\ was omitted in sect. 6 of ref. [\GR].) Once eq. \seex\ is imposed,
the number of unknowns and equations for the $Y$ coordinates
matches up properly.

\chapter{\bf Discussion}

This work has explored the noncompact $O(d,d)$ group that appears in
toroidal compactification of oriented closed bosonic strings as well as
the $O(d,d+n)$ generalization that is required for the heterotic string.
In sections 2 and 4 we showed, using methods of dimensional reduction,
that these noncompact groups are exact symmetries of the
(classical) low-energy effective field theory that is obtained
when one truncates the dependence on the internal coordinates
$y^{\alpha}$ keeping zero modes only.

In section 5 we explored noncompact symmetries from the world-sheet viewpoint,
extending the analysis of previous authors [\NSW,\CFG,\DUFFB]
to a somewhat more general
setting. We found that the classical string dynamics that results from
toroidal compactification and zero-mode truncation is also described by
equations of motion that can be written in a manifestly
$O(d,d+n)$-invariant form. Only global boundary conditions break the
symmetry to the discrete subgroup $O(d,d+n,Z)$. Therefore the
moduli space that arises in toroidal string compactification
is given by the $O(d,d+n)$ group manifold modded out by $O(d,d+n,Z)$
as well as by the maximal compact subgroup $O(d)\times O(d+n)$.

Logically, the analysis of section 5 should perhaps come first, since it
describes the noncompact symmetry at tree-level of the $\sigma$ model,
{\it i.e.}, to leading order in the $\alpha'$ expansion. The low-energy
effective field theory analysis of sections 2 and 4 corresponds
to the requirement of conformal invariance of the sigma model at
the one-loop order [\FRADKIN,\CFMP].
In particular, at this order the sigma model
action must be modified to include a term coupling the dilaton to the
world-sheet curvature [\FRADKIN].
We have not investigated the higher-loop corrections, which generate
additional higher-dimension terms in the field  equations of the
massless fields. They could in principle be generated by enforcing
conformal symmetry of the world-sheet action to higher orders in
$\alpha'$. It seems very plausible that the noncompact symmetries
would continue to hold for them as well. For example, a strong case
could probably be made by using formal path-integral manipulations along
the lines described by Fradkin and Tseytlin [\FT].
In fact, some evidence that the $O(d,d)$ symmetry is present at the
two-loop order has been presented by Panvel [\PANVEL], and more general
arguments have been advanced in refs. [\MO,\GR].

One result that seems interesting
to us is that the need for Chern--Simons terms in the $H_{\mu\nu\rho}$
field strength was deduced from purely bosonic considerations.
One wonders whether two-loop conformal invariance implies
the necessity of Lorentz Chern--Simons terms, again from
purely bosonic considerations.

The noncompact symmetries transform the moduli fields in complicated
nonlinear ways. In section 4 we reviewed techniques (well-known from
previous supergravity studies) for realizing these symmetries linearly.
Two distinct constructions to achieve this were presented. The first one was
a bit of a surprise, whereas the second was the standard coset
construction in which one introduces auxiliary scalar fields to fill out
the adjoint representation of the noncompact group and then compensates
by introducing a
local gauge symmetry corresponding to the maximal compact subgroup. This is
implemented using a generalized `vielbein' formalism, which we saw gives
rise to a better understanding of some of the otherwise mysterious
matrices that appear.

In the special (but physically interesting) case $D=4$, it is well known
that there is an additional $SU(1,1)$ or $SL(2,R)$ symmetry of the
low-energy effective field theory. The special feature of four dimensions
is that by making a duality transformation it is possible to replace the
antisymmetric tensor $B_{\mu\nu}$ by a scalar field, usually called the
`axion'. The axion and dilaton together then magically parametrize the
coset space $SL(2,R)/SO(2)$. The full $SL(2,R)$ symmetry in the presence
of vector fields ${\cal A}_{\mu}$ cannot be realized on the action, but
can be understood in terms of the classical field equations [\CSF]. (This
involves duality transformations of the vector fields.) The way this
works is rather analogous to the way the $O(d,d+n)$ symmetry is realized
on the world sheet. There too the symmetry could only be made manifest
for the field equations. Despite these common features, the $SL(2,R)$
symmetry appears to be of a qualitatively different character than the
$O(d,d+n)$ symmetry. The evidence for this is that it is apparently
impossible to realize it on the classical string equations of section 5.
However, this question still deserves further investigation.

Veneziano and collaborators [\V] have considered the $O(d,d)$ effective
theory with background fields $B_{\alpha\beta}(t)$ and
$G_{\alpha\beta}(t)$ depending on ``time'' only.  The action they
arrive at is
$$	S = \int dt e^{-\phi} [\Lambda + (\dot\phi)^2 + {1\over 8} tr (\dot M
\eta \dot M \eta)]\,\, ,\eqn\sfa$$
where $\Lambda$ is a cosmological constant proportional to $D-D_{crit}$.
Solutions to the classical field equations
obtained from \sfa\ describe spatially homogeneous cosmological models.
They exploit the global symmetry of the theory to generate new solutions
that would have been difficult to discover by other
methods.  Transformations  involving dimensions
that are not compactified should correspond to exact symmetries, even
in the string case.  However, for uncompactified dimensions, one has to
mod out $O(d,d)$ by $GL(d,R)$ and by constant shifts in $B$. The
resulting coset was identified by Sen [\HS] with $O(d) \times O(d)/
O(d)$.  Sen has considered more general models of this type in
order to obtain new black hole and black string solutions [\SEN].  His
techniques appear to be quite powerful.

The emphasis in our work has been to understand the common origin of
noncompact groups in string theory and field theory, both as symmetry
groups of low energy effective action and for the characterization of
string theory moduli spaces.  For toroidally compactified dimensions,
the only case studied in detail, this has been achieved.  Clearly, it
would be desirable to explore extensions and generalizations appropriate
to other internal spaces $K$. For example, we know from the work of
Seiberg that K3 compactification of the heterotic string should give a
six-dimensional theory with a $O(20,4)$ coset structure [\SEIBERG].
(This is remarkably similar to what one gets from $T^4$ compactification,
though the two cases do seem to be somewhat different [\GIVEON].)

Calabi--Yau spaces are of particular interest in string theory, since
in the context of heterotic
string compactification they can lead to many realistic features [\CHSW].
In addition to the $SU(1,1)/U(1)$ associated with the axion-dilaton
system, the moduli space of the heterotic string compactified on
a Calabi--Yau manifold
\foot{For a description of the geometry of Calabi--Yau moduli space see
ref. [\Xenia] and references therein. This subject has been very active
in recent years, and we will not attempt to give a complete
set of references here.}
consists of two factors, ${\cal M}_{11} \times {\cal M}_{21}$,
where ${\cal M}_{11}$
corresponds to K\"ahler form deformations and has complex dimension
$h_{11}$, while ${\cal M}_{21}$ describes complex structure deformations and
has complex dimension $h_{21}$. (This factorization was established in
refs. [\CFGB,\DKL].)  The integers $h_{11}$ and $h_{21}$ are
Hodge numbers of the Calabi--Yau space.  In general, each factor should
have a discrete symmetry group analogous to the $O(d,d + n, Z)$ of
toroidal compactification.  It is quite difficult to compute the
groups for specific examples, but it is known that they must be
subgroups of $Sp(2b_{11} + 2, Z)$ and $Sp (2b_{21} + 2, Z)$,
respectively. (One specific CY example has been worked out in detail in ref.
[\CXGP]. An orbifold example is given in ref. [\FFS].)

The spaces ${\cal M}_{11}$ and ${\cal M}_{21}$
are themselves K\"ahler manifolds of a
special type for which the K\"ahler potential can be derived from a
holomorphic prepotential [\FERRARA].
Homogeneous spaces of this type have been
classified. Presumably, at least in certain cases, Calabi--Yau moduli
spaces are given by such homogeneous spaces modded out by the discrete
group. Whether or not this is the general case (we do not know), it may
be interesting to try to classify Calabi--Yau spaces whose moduli spaces
are of this type. For this
class, the techniques described in this paper for tori should have the most
straightforward generalizations.

In conclusion, there is much more still to be learned by pursuing the
study of noncompact groups of the type described here.  In string theory
they are broken to discrete subgroups.  These subgroups are, in fact,
``discrete gauge symmetries,'' [\HUET] which means that they should be quite
robust, surviving the plethora of phenomena that typically break global
symmetries.  By thinking hard about them, it may be possible to draw
some very powerful general conclusions about compactified dimensions,
as well as the implications for physical four-dimensional spacetime.

\ack

One of us (J.M.) would like to acknowledge the gracious hospitality of the
Theoretical Particle Physics group at Caltech. We are grateful to G.
Veneziano for reading the manuscript and suggesting improvements.

\refout
\bye